\documentclass[pdflatex,sn-mathphys-num]{sn-jnl}

\usepackage{graphicx}%
\usepackage{multirow}%
\usepackage{amsmath,amssymb,amsfonts}%
\usepackage{amsthm}%
\usepackage{mathrsfs}%
\usepackage[title]{appendix}%
\usepackage{xcolor}%
\usepackage{textcomp}%
\usepackage{manyfoot}%
\usepackage{booktabs}%
\usepackage{algorithm}%
\usepackage{algorithmicx}%
\usepackage{algpseudocode}%
\usepackage{listings}%
\usepackage{url}
\usepackage[utf8]{inputenc}
\usepackage[T1]{fontenc}
\usepackage{lmodern}

\theoremstyle{thmstyleone}%
\theoremstyle{thmstyletwo}%

\theoremstyle{thmstylethree}%

\begin{document}

\title{Machine Learning Multiscale Interactions}



\author[1,2,3]{\fnm{Àlex} \sur{Solé}}\email{jaume.alexandre.sole@upc.edu}

\author[3]{\fnm{Sergio} \sur{Suárez-Dou}}\email{ sergio.suarezdou@uni.lu}

\author[1]{\fnm{Albert} \sur{Mosella-Montoro}}\email{albert.mosella@upc.edu}

\author*[2]{\fnm{Silvia} \sur{Gómez-Coca}}\email{silvia.gomez@qi.ub.edu}

\author*[2]{\fnm{Eliseo} \sur{Ruiz}}\email{eliseo.ruiz@qi.ub.edu}

\author*[3]{\fnm{Alexandre} \sur{Tkatchenko}}\email{alexandre.tkatchenko@uni.lu}

\author*[1]{\fnm{Javier} \sur{Ruiz-Hidalgo}}\email{j.ruiz@upc.edu}

\affil[1]{\orgdiv{Image Processing Group -- Signal Theory and Communications Department}, \orgname{Universitat Politècnica de Catalunya}, \orgaddress{\city{Barcelona},  \country{Spain}}}

\affil[2]{\orgdiv{Inorganic and Organic Chemistry Department and Institute of Theoretical and Computational Chemistry}, \orgname{Universitat de Barcelona}, \orgaddress{\city{Barcelona},  \country{Spain}}}

\affil[3]{\orgdiv{Department of Physics and Materials Science}, \orgname{University of
Luxembourg}, \orgaddress{\city{Luxembourg City}, \country{Luxembourg}}}


\abstract{Realistic physical systems are characterised by emergent interactions across multiple length and time scales, posing a significant challenge for predictive machine learning (ML) models. Most scientific ML models focus on a narrow range of interactions. While machine learning force fields (MLFFs) offer near-quantum accuracy, the ubiquitous message-passing layers miss long-range many-body effects. Here we introduce the Multiscale Structural Ensemble (MuSE), a hierarchical model that uses Soft Coarse-Graining Pooling
to construct coarse representations from smooth fractional assignments of atoms to coarse nodes, enabling MLFF modules to operate across multiple scales. MuSE is architecture-agnostic and coupled with SO3krates, MACE, and PaiNN MLFFs for both molecules and materials. 
We demonstrate the power of MuSE through Hessian-based benchmarks, folding trajectories for biomolecules, and energy profiles in molecule--graphene nanostructures, where MuSE accurately captures quantum-mechanical interactions at relevant scales---unlike other recent long-range ML models.
}

\maketitle

\section*{Introduction}

Many scientific learning problems are governed by interactions and correlations that emerge across multiple spatial and temporal resolutions. In structured data, target observables may depend not only on local neighbourhoods, but also on mesoscopic organisation, long-range couplings, and global structural information. Similar multiscale dependencies appear in many types of data, including molecular and materials graphs~\cite{reiser2022gnnchem}, point clouds and geometric perception data~\cite{qi2017pointnet, qi2017pointnet++}, multiscale scientific simulations~\cite{sanderse2025multiscale}, and spatiotemporal relational data~\cite{capone2025stgnn}. Across these domains, the relevant signal is often distributed across several scales rather than confined to a single local receptive field. A central challenge for machine-learning models is therefore to construct representations that preserve fine-scale detail whilst remaining sensitive to broader and collective structure.

Machine-learning force fields (MLFFs) provide a particularly demanding instance of this general problem. Trained on quantum-mechanical data, they enable the calculation of energies and forces at a computational cost adequate for large-scale molecular simulations~\cite{schnet,unke2021spookynet,MD22,kabylda2023efficient}. Most contemporary MLFFs utilise graph neural networks (GNNs)~\cite{gilmer2017neural}, in which atoms are represented as nodes, geometric relationships are represented by edges, and atomic features are updated via message-passing. Message-passing denotes the iterative exchange and aggregation of information between neighbouring atoms, enabling the model to construct local many-body representations of the atomic environment. This approach has demonstrated success in predicting molecular and materials properties~\cite{cartnet,conformer,potnet,matformer} as well as interatomic potentials~\cite{schnet,gasteiger_dimenet_2020,gasteiger_dimenetpp_2020,nequip,allegro,schutt2021painn,mace,frank2022so3krates,so3krates2}. Modern architectures further incorporate the symmetries intrinsic to atomistic physics: energies remain invariant under translation, rotation, and permutation of identical atoms, while forces transform equivariantly under rotations. Although this framework has achieved high accuracy across a wide range of molecular and materials benchmarks, its locality remains a fundamental limitation: information is typically restricted by a finite cutoff and a small number of message-passing layers. Consequently, slowly decaying and collective contributions, such as electrostatics, polarisation, dispersion, conformational coupling, and elastic responses, are not readily recovered when they lie beyond the local atomic neighbourhood. Increasing the cut-off radius densifies the graph, whereas adding more message-passing layers can result in oversmoothing or oversquashing of propagated information~\cite{demystify_gat_oversmooth2023,dropedge2020,pairnorm2020}.

The main challenge is thus not only to enable each atom to receive information from more distant parts of the structure, but also to recover non-local couplings while preserving the atomistic resolution and smoothness essential for molecular dynamics. Multiple strategies have been created to address this challenge. Some approaches modify the graph topology through incorporating auxiliary neighbourhoods, angular or feature-space graphs, or sparse global descriptors~\cite{gasteiger2021gemnet,alignn,frank2022so3krates,PRISM,kabylda2023efficient,fastattention_frank}. Other methods introduce explicit physical contributions, such as learned electrostatics or analytic long-range corrections~\cite{unke2021spookynet,GEMS,so3lr}. Additional techniques employ reciprocal-space, Ewald-like, latent, or virtual-node communication modules atop short-range neural potentials~\cite{ewald_sum,ewald_sum_2,range}. Related methods utilise learned charges or global molecular representations to transmit information beyond local neighbourhoods~\cite{ko2021fourth,king2025chargeslongrange,zhou2023unimol,mendezlucio2024mole}. While these strategies underline the necessity of non-local information, they also reveal a modelling tension: long-range effects are frequently included through analytic physical forms or compressed global channels, rather than through a hierarchy that retains distance-resolved many-body structure. We show that current approaches struggle to represent multiscale many-body interactions in large molecules.

Instead, here we introduce the Multiscale Structural Ensemble (MuSE), a model-agnostic architecture that augments a base MLFF with a learned hierarchy of coarse representations. This approach is analogous to multiscale representation learning in point-cloud processing and geometric computer vision, where progressively coarser neighbourhoods combine local detail with broader geometric context~\cite{spotr,qi2017pointnet,qi2017pointnet++,wang2019ddgcn,kpconv,hu2020randlanet,zhao2021pointtransformer,landrieu2018spg,MOSELLAMONTORO2021}. The central component is Soft Coarse-Graining Pooling (SCGP), a smooth, symmetry-compatible pooling operation that maps atoms onto coarse nodes through fractional rather than hard assignments. The key feature of this pooling is that an atom is not assigned exclusively to a single coarse bead. Instead, each atom can contribute with different percentages to several neighbouring coarse nodes. Because these fractional assignments vary continuously with the molecular geometry, SCGP enables recursive multiscale pooling while preserving the differentiability required for force-field applications. Information from each scale is then propagated back to the atomistic resolution, where the final energy and forces are predicted. As a result, no reference energies or forces are required for the coarse beads, which serve only as internal latent variables to transmit multiscale information. MuSE is independent of the choice of message-passing architecture and can be coupled with widely used models such as SO3krates~\cite{so3krates2}, MACE~\cite{mace}, and PaiNN~\cite{schutt2021painn}. It is also applicable to both finite and periodic systems.

MuSE was trained and evaluated on molecular and materials benchmarks. It improves energy and force accuracy, recovers long-range Hessian structure more accurately than existing long-range strategies, and preserves stable molecular dynamics. Energy error reductions of up to 41.3\% and force error reductions of up to 26.5\% are achieved, with consistent performance gains on MD22 \cite{MD22} and TEA2023 \cite{crashtest}, as well as improvements across SO3krates, MACE, and PaiNN backbones. Comparisons with state-of-the-art long-range models, Ewald-style~\cite{ewald_sum}, RANGE~\cite{range}, and SO3LR~\cite{so3lr} methods indicate that MuSE provides the strongest overall agreement with DFT long-range couplings. Hence, MuSE represents the first model-agnostic multiscale coarse-graining framework for equivariant MLFFs.

\section*{Results \& Discussion}

\subsection*{Building a hierarchical multiscale representation with MuSE}

\begin{figure}[!htb]
    \centering
    \includegraphics[width=\linewidth]{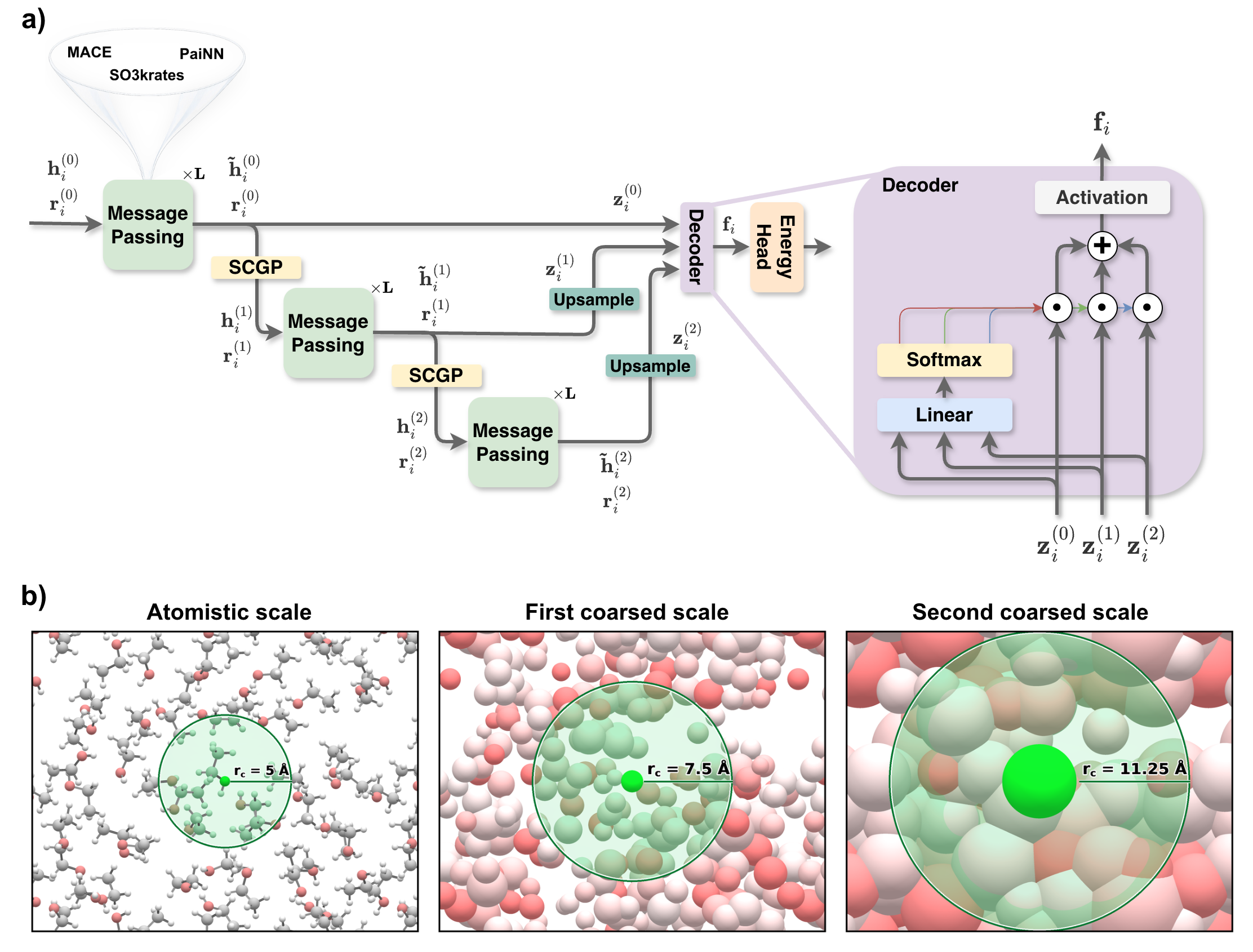}
    \caption{MuSE architecture and multiscale receptive fields. \textbf{a)} Starting from the atomistic graph, a base MLFF block updates node embeddings at each scale. Soft Coarse-Graining Pooling (SCGP) constructs coarser graphs through smooth many-to-one aggregation, after which message-passing is applied on the pooled representations. Coarse embeddings are upsampled with the stored SCGP weights to obtain atom-wise multiscale representations $\mathbf{z}_i^{(s)}$. The decoder assigns atom-specific weights to each scale. It then combines the corresponding representations, applies an activation function, and passes the latent $\mathbf{f}_i$ to the energy head. \textbf{b)} Illustration of the hierarchy: as the number of nodes decreases, the cut-off radius increases, so each coarse node summarises a broader structural region.}
    \label{fig:architecture}
\end{figure}

A multiscale representation provides a direct way to extend the receptive field without abandoning atomistic predictions. Fine-scale features can be compressed into coarser graphs, processed over wider neighbourhoods, and returned to the atomic resolution before the energy prediction. The central requirement is that this compression and reconstruction remain smooth functions of the atomic coordinates. In geometric learning, Farthest Point Sampling (FPS)~\cite{FPS} and voxelisation~\cite{voxel,MOSELLAMONTORO2021} compress spatial data efficiently, but they are less suitable for atomistic force fields. FPS keeps selected representatives, so the remaining atoms do not contribute continuously to the pooled representation. Voxel-based schemes impose discrete spatial partitions, which can introduce frame dependence or discontinuities. These effects are problematic in molecular dynamics, where the potential-energy surface must vary smoothly to yield conservative forces.

MuSE addresses this limitation with a smooth hierarchy of coarse graph representations. SCGP assigns neighbouring atoms fractionally to coarse nodes through continuous many-to-one weights, so that coarse variables remain differentiable functions of the atomic coordinates. As shown in Fig.~\ref{fig:architecture}a, message-passing first refines atomistic features. SCGP then builds coarser graphs, on which the same type of message-passing block is applied. This operation changes the graph resolution, as illustrated in Fig.~\ref{fig:architecture}b. The number of nodes decreases and the cut-off radius increases. Each coarse node inherits local information from the lower representation through its geometry-dependent weights. Unlike an isotropic bead, which would represent a region only through an averaged position or density, a MuSE coarse node retains structured information about the spatial arrangement and chemical identities of its contributing atoms. It therefore acts as an anisotropic descriptor of a molecular domain. Interactions between these descriptors at the next scale provide an effective way to represent many-body correlations: the state of each coarse node already encodes a collective atomic environment, and message-passing between coarse nodes learns how such environments interact over larger distances. The resulting embeddings are then upsampled to atomistic resolution with the stored SCGP weights and fused by an atom-wise gated decoder before the energy head. Coarse nodes thus propagate collective structural information across scales while the atomistic branch remains available for local chemistry. MuSE thereby extends the effective receptive field through a physics-inspired coarse-graining mechanism, while retaining atomistic resolution and symmetry-compatible force prediction. A full description of the architecture is provided in the Methods and Supplementary Information Section~S1.

\subsection*{Multiscale modelling recovers long-range couplings and dynamics}

Gas-phase folding of alanine peptides provides a stringent benchmark for MLFFs, as their conformational preferences arise from a balance between local bonded interactions and long-range collective many-body effects~\cite{AlexPRL2011}. Helical stabilisation is governed not only by backbone torsions, but also by non-local electrostatics, dispersion interactions, hydrogen-bond networks, and correlated motions extending across the full molecular length. Recent MLFFs, including the MACE family~\cite{mace} and SpookyNet~\cite{GEMS}, yield reproducible folding trends. However, their finite locality or simplified treatment of non-local interactions can limit their ability to recover the full long-range coupling structure underlying conformational stability.

We compared MuSE with representative strategies for extending the effective range of machine-learned force fields. Plain SO3krates is the reference model without a long-range correction. It uses a 5~\AA{} radius graph and three message-passing layers. Two controls keep the same local modelling framework but increase the amount of information that can be exchanged. SO3krates (10~\AA{}) widens the radius graph without changing the number of trainable parameters. SO3krates (9 layers) instead deepens the network and matches the parameter count of SO3krates-MuSE. The long-range variants then add explicit non-local communication: SO3krates-Ewald introduces reciprocal-space communication~\cite{ewald_sum}, RANGE~\cite{range} transmits information through learned global encodings, and SO3krates-MuSE uses the multiscale hierarchy described above. All models in Table~\ref{tab:ablations_ef} were trained on the alanine peptide subset from GEMS~\cite{GEMS}, comprising polyalanine chains of increasing length.

\begin{table}[!htb]
\centering
\caption{Energy and force MAE, together with the number of trainable parameters, for SO3krates-based variants on the alanine peptide subset from GEMS. Values are reported as mean $\pm$ standard deviation across four runs. Lower values are better for energy and force. Values in parentheses in the MAE columns denote the relative improvement with respect to SO3krates. Bold marks the best MAE and underlining the second-best MAE for each metric.}
\label{tab:ablations_ef}
\begingroup
\scriptsize
\setlength{\tabcolsep}{3pt}
\begin{tabular*}{\textwidth}{@{\extracolsep\fill}lccc}
\toprule
& \textit{E} $\downarrow$ & \textit{F} $\downarrow$ & \# Params \\
\textbf{Method} & $(kcal/mol)$ & (kcal/mol/\AA) & $(M)$ \\
\midrule
SO3krates            & 1.748 $\pm$ 0.095 & 0.431 $\pm$ 0.014 & 0.43 \\
SO3krates (10~\AA{}) & 1.066 $\pm$ 0.104 (+39.0\%) & 0.379 $\pm$ 0.022 (+12.0\%) & 0.43 \\
SO3krates (9 layers) & 1.272 $\pm$ 0.125 (+27.2\%) & \underline{0.358 $\pm$ 0.016 (+16.9\%)} & 1.22 \\
SO3krates-Ewald      & 1.160 $\pm$ 0.068 (+33.6\%) & 0.371 $\pm$ 0.019 (+13.8\%) & 0.64 \\
SO3krates-RANGE      & \underline{1.104 $\pm$ 0.062 (+36.8\%)} & \textbf{0.342 $\pm$ 0.013 (+20.7\%)} & 2.80 \\
SO3krates-MuSE       & \textbf{1.026 $\pm$ 0.044 (+41.3\%)} & 0.382 $\pm$ 0.010 (+11.2\%) & 1.22 \\
\bottomrule
\end{tabular*}
\endgroup
\end{table}

Table~\ref{tab:ablations_ef} shows that adding a long-range correction or enlarging the radius graph improves the scalar errors relative to plain SO3krates. The energy errors improve across these variants, and the force errors become broadly similar once some form of extended-range information is included. The main difference, therefore, lies between the corrected models and the uncorrected SO3krates baseline, rather than among the long-range strategies themselves. These standard metrics alone do not show where the improvement comes from. The gains can arise from more trainable parameters, additional interaction terms, a more detailed local environment, or a more accurate description of long-range couplings. Since these changes are mixed across the variants, with energies and forces alone, we cannot identify which effect is responsible. This motivates analysing the distance dependence of the learned couplings directly.

The Hessian of the potential energy with respect to atomic displacements provides this test. As the second-order term in the expansion of the potential-energy surface, it characterises the linear force response on atom $i$ to a displacement of an atom $j$ for a given geometry. It therefore allows us to compare second-order couplings across models with the electronic-structure reference. In simple pairwise potentials, off-diagonal Hessian elements decay rapidly with interatomic distance. By contrast, DFT calculations can retain longer-range contributions through electrostatics and many-body effects. We assess whether MLFF models reproduce the distance dependence of these couplings. This analysis complements standard energy and force metrics. For this Hessian comparison, we also included SO3LR~\cite{so3lr}, a foundational long-range model that incorporates electrostatics and dispersion through universal pairwise terms. We used the available pretrained SO3LR model for this diagnostic. The reference \textit{ab initio} Hessian at the PBE0~\cite{pbe0}+MBD~\cite{mbd} level displays non-negligible long-range structure. It therefore provides a stringent benchmark for the local linear response. We used four representative conformers of the Ace-Ala\textsubscript{15}-NMe folding profile~\cite{hessians_sergio}: an $\alpha$-helical conformer, a $3_{10}$-helical conformer, a nearly unfolded conformer, and an unfolded conformer. These conformers span pairwise distances from 25~\AA{} to 60~\AA{}.

\begin{figure}[p]
    \centering
    \includegraphics[width=\textwidth,height=0.889\textheight,keepaspectratio]{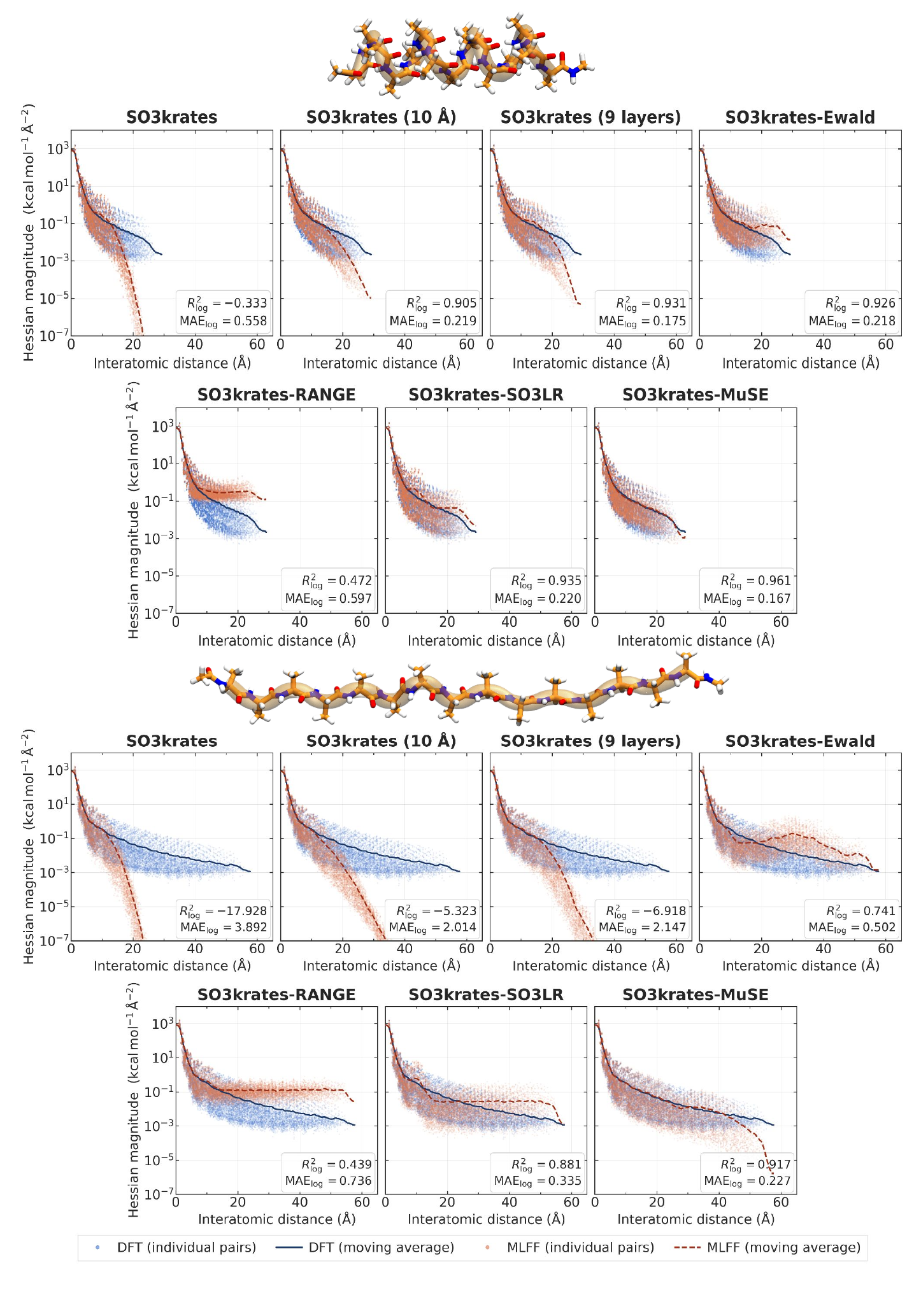}
    \caption{Comparison of Hessian interaction--distance profiles for different long-range strategies on representative Ace-Ala$_{15}$-NMe conformers. The top row corresponds to the $\alpha$-helical conformer and the bottom row to the unfolded conformer. For each method, predicted Hessian elements (orange) are plotted as a function of interatomic distance and compared with the corresponding DFT reference (blue). Scatter points denote individual Hessian entries, while the overlaid curves show the corresponding moving averages.}
    \label{fig:hessian_comparison}
\end{figure}

Hessian interaction--distance profiles across methods are illustrated in Fig.~\ref{fig:hessian_comparison}. For local SO3krates models, increasing either the cutoff or the number of message-passing layers extends the nominal receptive field, but does not lead to a proportional improvement in the recovery of long-range couplings. In all cases, the predicted profiles begin to deviate markedly from the DFT reference at substantially shorter distances than suggested by their formal interatomic distance range. This shows that simply enlarging the local graph or deepening the network is insufficient to recover the physically correct distance dependence of the Hessian, consistent with previous observations regarding the recovery of long-range atomic interactions~\cite{AtomicInteractions}.

The other explicitly long-range models do not fully reproduce the reference behaviour, as shown in Fig.~\ref{fig:hessian_comparison}. Although Ewald, RANGE, and SO3LR provide formal non-local communication, their Hessian interaction--distance profiles display clear qualitative artefacts relative to the DFT reference, including oscillatory behaviour, anomalous intermediate-distance structure, and unphysical flattening at long range. These artefacts point to a limitation shared by such mechanisms: they extend the communication range without preserving a distance-resolved many-body representation of the interaction. In Ewald-like approaches, non-local communication is mediated through a global spectral representation. This is well suited to smooth long-range fields, but it does not explicitly retain a real-space hierarchy of many-body couplings. As a result, the Ewald profile in Fig.~\ref{fig:hessian_comparison} shows that the global channel may introduce oscillatory or delocalised contributions. Extended communication therefore does not necessarily imply that the resulting Hessian elements recover the correct real-space attenuation as a function of interatomic distance. In global-encoding approaches such as RANGE, information is compressed into global virtual-node representations and then broadcast to all atoms. The RANGE profiles in Fig.~\ref{fig:hessian_comparison} are consistent with this low-rank, mean-field communication bottleneck, which reduces sensitivity to pair-specific distance variations. In analytic pairwise approaches such as SO3LR, the imposed long-range channel provides a useful physical prior. However, the SO3LR profiles in Fig.~\ref{fig:hessian_comparison} indicate that its functional form may not fully capture environment-dependent screening, polarisation, and conformational many-body effects that contribute to the Hessian. Thus, the comparison shows that extending the interaction range in itself is not enough. The decay of interatomic couplings must also be described correctly.

By contrast, Fig.~\ref{fig:hessian_comparison} shows that MuSE most closely reproduces the DFT Hessian profile over a broad and physically meaningful distance range. Its predictions remain in quantitative agreement with the reference up to approximately 45~\AA, far beyond the regime accessible to the other approaches. The key advantage of MuSE is therefore not merely a larger receptive field, but its ability to encode the multiscale physics governing long-range force propagation. The coarse-to-fine structure of MuSE accounts for this distinction. As illustrated by the cutoff and depth variants in Fig.~\ref{fig:hessian_comparison}, increasing the cutoff or the number of message-passing layers still keeps communication at a single atomistic resolution. Long-range information must therefore pass through many local updates, where it can become diluted, oversmoothed, or mixed with irrelevant local signals. In MuSE, distant atomic regions can communicate through coarse nodes that summarise larger structural domains. This shortens the effective communication path while retaining the fine-scale atomistic branch required to resolve local chemical detail, consistent with the closer agreement between the MuSE and DFT moving averages in Fig.~\ref{fig:hessian_comparison}. Results for the remaining conformers and the numerical comparison between methods are provided in Supplementary Information, Section~S2. The change in the Hessian profile with the number of MuSE scales is examined in Supplementary Information, Section~S3.

\begin{table}[!htb]
\centering
\caption{Computational statistics for the different long-range strategies, reported as per-step MD wall-clock time and peak GPU memory. Values are reported as mean $\pm$ standard deviation. Lower values are better for both metrics. Values in parentheses denote the relative change with respect to SO3krates. Bold marks the best result and underlining the second-best result for each metric.}
\label{tab:hessian_cost}
\begingroup
\scriptsize
\setlength{\tabcolsep}{3pt}
\begin{tabular*}{\textwidth}{@{\extracolsep\fill}lcc}
\toprule
& MD time step $\downarrow$ & GPU peak memory $\downarrow$ \\
\textbf{Method} & $(\mathrm{ms})$ & $(\mathrm{MB})$ \\
\midrule
SO3krates & \textbf{10.40 $\pm$ 0.49} & \textbf{95.05 $\pm$ 10.36} \\
SO3krates (10~\AA{}) & \underline{10.69 $\pm$ 0.71} (-2.8\%) & 232.28 $\pm$ 68.69 (-144.4\%) \\
SO3krates (9 layers) & 31.55 $\pm$ 13.54 (-203.4\%) & 246.05 $\pm$ 25.52 (-158.9\%) \\
SO3krates-Ewald & 13.12 $\pm$ 0.42 (-26.2\%) & 156.88 $\pm$ 7.40 (-65.1\%) \\
SO3krates-RANGE & 15.86 $\pm$ 0.57 (-52.5\%) & \underline{115.56 $\pm$ 5.23} (-21.6\%) \\
SO3krates-MuSE & 25.29 $\pm$ 0.59 (-143.2\%) & 174.82 $\pm$ 26.83 (-83.9\%) \\
\bottomrule
\end{tabular*}
\endgroup
\end{table}

Recovering long-range couplings is only useful if the resulting model remains computationally practical for molecular dynamics. The computational statistics are summarised in Table~\ref{tab:hessian_cost}. The baseline SO3krates model remains the least expensive configuration, as expected. Increasing the cutoff to 10~\AA{} yields a similar MD step time, but substantially increases peak GPU memory usage. Increasing the depth to nine layers is more costly in both time and memory. The Ewald-based model introduces a moderate overhead relative to SO3krates, whereas RANGE is comparatively memory-efficient. MuSE occupies an intermediate regime: it is more expensive than the baseline but less memory-demanding than the uniformly deeper SO3krates model. Thus, MuSE offers a favourable accuracy--efficiency trade-off, improving the recovery of long-range couplings while remaining computationally practical for molecular dynamics.

\begin{figure}[!htb]
    \centering
    \includegraphics[width=\linewidth]{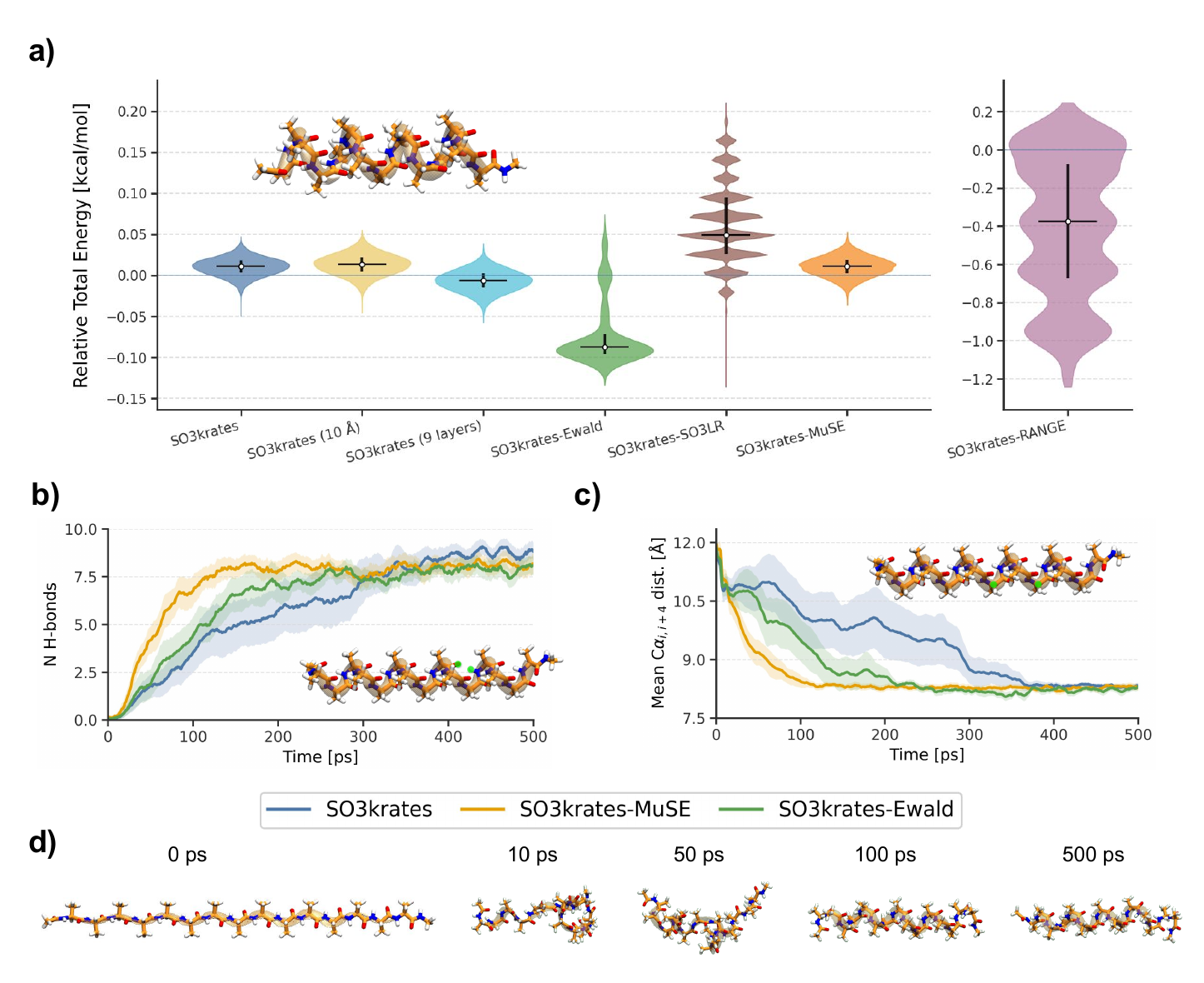}
    \caption{Molecular-dynamics stability and folding signatures for Ala$_{15}$. \textbf{a,} Relative total-energy distributions from 1~ns NVE simulations initialised from the $\alpha$-helical Ace-Ala$_{15}$-NMe structure. Relative total energy is defined as the instantaneous total energy minus the initial total energy. White markers show the mean, black horizontal bars the median, and black vertical bars the interquartile range; the horizontal blue line marks zero relative total energy. \textbf{b,} Number of backbone hydrogen bonds during folding simulations. Lines show the mean and shaded regions indicate the standard error of the mean (SEM). \textbf{c,} Mean $C_{\alpha,i}$--$C_{\alpha,i+4}$ distance during the same folding simulations; shaded regions indicate the SEM. \textbf{d,} Representative conformations from a SO3krates-MuSE folding trajectory at selected simulation times. Statistical comparisons with SO3krates were performed using Student's $t$-tests and a significance threshold of $p < 0.05$. SO3krates-MuSE is the only method that differs significantly from SO3krates in both folding metrics. For \textbf{b}, $p = 0.031$ for SO3krates-MuSE and $p = 0.357$ for SO3krates-Ewald. For \textbf{c}, $p = 0.028$ for SO3krates-MuSE and $p = 0.161$ for SO3krates-Ewald.}
    \label{fig:vdos-alanine}
\end{figure}

Beyond computational practicality, a useful force field must also yield stable molecular dynamics. This requires a sufficiently smooth potential-energy surface to support energy-conserving propagation. To assess this, we performed 1~ns microcanonical dynamics simulations and monitored the conservation of total energy (Fig.~\ref{fig:vdos-alanine}a). The semi-local models, including SO3krates, remain effectively energy-conserving, consistent with their smooth potentials. SO3krates-SO3LR and SO3krates-Ewald both show clear departures, with broader distributions and systematic offsets towards higher and lower relative energies, respectively. RANGE performs worst, showing clear deviations from conservative dynamics, consistent with a high level of numerical noise. Notably, MuSE combines physically accurate long-range interactions with the smoothness of semi-local models, thereby preserving stable and energy-conserving dynamics. 

To assess whether improved long-range communication affects folding dynamics, we performed NVT molecular dynamics starting from the unfolded Ace-Ala$_{15}$-NMe structure and analysed 10 replicas per method that undergo a direct transition from a disordered coil to a helical configuration. The hydrogen-bond trajectories show that adding a long-range module accelerates the formation of helical order (Fig.~\ref{fig:vdos-alanine}b). SO3krates forms hydrogen bonds more gradually, whereas SO3krates-Ewald and SO3krates-MuSE reach a larger hydrogen-bond count earlier in the simulation. This effect is accompanied by a faster decrease in the mean $C_{\alpha,i}$--$C_{\alpha,i+4}$ distance (Fig.~\ref{fig:vdos-alanine}c), indicating earlier compaction of local helical turns. The trend is observed for both SO3krates-Ewald and SO3krates-MuSE, but is more pronounced for MuSE, which reaches a compact helical geometry rapidly and stabilises a hydrogen-bond network characteristic of the folded state. The differences between SO3krates-MuSE and SO3krates were statistically significant for both folding metrics ($p < 0.05$). These trends indicate that global communication introduced by long-range modules provides a more effective description of non-local intramolecular couplings, thereby steering the chain towards folded configurations consistent with the overall organisation of the peptide. In contrast, a strictly local model is less constrained by the global conformation during the early stages of collapse, as the effective forces are dominated by short-range atomic environments. These results complement the Hessian analysis by showing that long-range interactions are not only important for recovering extended force couplings, but also have a direct impact on molecular dynamics~\cite{AlexPRL2011, Rossi2013Impact}.

\subsection*{Accuracy evaluation in diverse benchmarks}

We first assess whether MuSE improves the description of chemically diverse systems in which non-local effects arise from different physical origins. To this end, we consider two complementary benchmarks: MD22~\cite{MD22} and TEA2023~\cite{crashtest}. MD22 contains flexible molecules, supramolecular assemblies, nucleic-acid fragments, and nanostructures with varying sizes, topologies, and interaction ranges. Within this benchmark, the double-walled nanotube is of particular interest. Its extended geometry produces large end-to-end separations, and its concentric structure gives rise to non-local correlations that are difficult to resolve with a strictly local atomic neighbourhood. TEA2023 provides a complementary condensed-phase setting. It includes a peptide, an organic adsorbate on graphene, and the hybrid organic--inorganic perovskite MAPbI$_3$. This benchmark allows us to test whether the multiscale representation remains useful under periodic boundary conditions, where molecule--surface interactions, electrostatics, and elastic responses can extend across the simulation cell.

Table~\ref{tab:combined_benchmarks_mae} summarises the absolute mean errors of energy and force for SO3krates and SO3krates-MuSE across MD22 and TEA2023. On MD22, MuSE improves both energy and force accuracy across all seven subsets, with average reductions of 25.5\% in energy MAE and 10.1\% in force MAE. The largest gains are obtained for the double-walled nanotube, where the energy MAE is reduced by 38.0\% and the force MAE by 22.5\%. The double-walled nanotube results are consistent with the idea that the multiscale hierarchy is especially beneficial when relevant interactions extend beyond the local receptive field. On TEA2023, MuSE also improves all energy and force errors, with average reductions of 20.6\% and 11.6\%, respectively. The strongest force improvement occurs for MAPbI$_3$, where the soft, polarisable inorganic framework is coupled to orientationally flexible molecular cations. Together, these benchmarks show that the multiscale pathway is not restricted to isolated finite molecules, but also transfers to periodic systems in which long-range electrostatic, elastic, and surface-mediated effects are important.

\begin{table}[!htb]
\centering
\caption{Energy (\textit{E}) and force (\textit{F}) MAE for the MD22 and TEA2023 benchmarks. Values are reported as mean $\pm$ std across four runs. Lower is better. Training-set sizes are given in parentheses, and the best value for each system is shown in \textbf{bold}. MuSE columns report the relative change with respect to the non-coarse SO3krates baseline in parentheses; positive indicates improvement. N-Ac-Phe-Ala$_5$-Lys denotes N-acetylphenylalanyl-pentaalanyl-lysine.}
\label{tab:combined_benchmarks_mae}
\tiny
\begin{tabular*}{\textwidth}{@{\extracolsep\fill}lcc|cc}
\toprule
& \multicolumn{2}{c}{\textbf{SO3krates}} & \multicolumn{2}{c}{\textbf{SO3krates-MuSE}} \\
\cmidrule(lr){2-3}\cmidrule(lr){4-5}
& \textit{E} & \textit{F} & \textit{E} & \textit{F} \\
\textbf{System} & $(\mathrm{kcal/mol})$ & $(\mathrm{kcal/mol/}$\AA$)$ & $(\mathrm{kcal/mol})$ & $(\mathrm{kcal/mol/}$\AA$)$ \\
\midrule
\multicolumn{5}{@{}l}{\textbf{MD22}} \\
\specialrule{0.08em}{0.3em}{0.15em}
Ace-Ala$_3$-NMe (6K) & 0.292 $\pm$ 0.028 & 0.287 $\pm$ 0.013 & \textbf{0.205 $\pm$ 0.015} (+29.7\%) & \textbf{0.267 $\pm$ 0.010} (+7.2\%) \\
DHA (8K) & 0.348 $\pm$ 0.026 & 0.289 $\pm$ 0.007 & \textbf{0.295 $\pm$ 0.036} (+15.1\%) & \textbf{0.266 $\pm$ 0.005} (+7.8\%) \\
Stachyose (8K) & 0.746 $\pm$ 0.260 & 0.467 $\pm$ 0.018 & \textbf{0.522 $\pm$ 0.168} (+30.0\%) & \textbf{0.418 $\pm$ 0.019} (+10.4\%) \\
AT--AT (3K) & 0.278 $\pm$ 0.049 & 0.261 $\pm$ 0.007 & \textbf{0.196 $\pm$ 0.016} (+29.3\%) & \textbf{0.240 $\pm$ 0.007} (+8.3\%) \\
AT--AT--CG--CG (2K) & 0.360 $\pm$ 0.049 & 0.344 $\pm$ 0.007 & \textbf{0.316 $\pm$ 0.027} (+12.5\%) & \textbf{0.312 $\pm$ 0.007} (+9.3\%) \\
Buckyball catcher (1K) & 0.404 $\pm$ 0.037 & 0.227 $\pm$ 0.010 & \textbf{0.309 $\pm$ 0.014} (+23.6\%) & \textbf{0.215 $\pm$ 0.008} (+5.3\%) \\
Double-walled (800) & 0.873 $\pm$ 0.123 & 0.560 $\pm$ 0.018 & \textbf{0.542 $\pm$ 0.011} (+38.0\%) & \textbf{0.434 $\pm$ 0.017} (+22.5\%) \\
\specialrule{0.12em}{0.5em}{0.2em}
\multicolumn{5}{@{}l}{\textbf{TEA2023}} \\
\specialrule{0.08em}{0.3em}{0.15em}
N-Ac-Phe-Ala$_5$-Lys (4K) & 0.625 $\pm$ 0.128 & 0.429 $\pm$ 0.013 & \textbf{0.477 $\pm$ 0.014} (+23.7\%) & \textbf{0.409 $\pm$ 0.011} (+4.8\%) \\
Naphthyridine/graphene (1K) & 0.279 $\pm$ 0.020 & 0.178 $\pm$ 0.006 & \textbf{0.242 $\pm$ 0.025} (+13.3\%) & \textbf{0.172 $\pm$ 0.007} (+3.4\%) \\
MAPbI$_3$ (500) & 0.702 $\pm$ 0.037 & 0.220 $\pm$ 0.013 & \textbf{0.527 $\pm$ 0.016} (+24.9\%) & \textbf{0.162 $\pm$ 0.007} (+26.5\%) \\
\bottomrule
\end{tabular*}
\end{table}

\begin{figure}[!htb]
    \centering
    \includegraphics[width=0.8\linewidth]{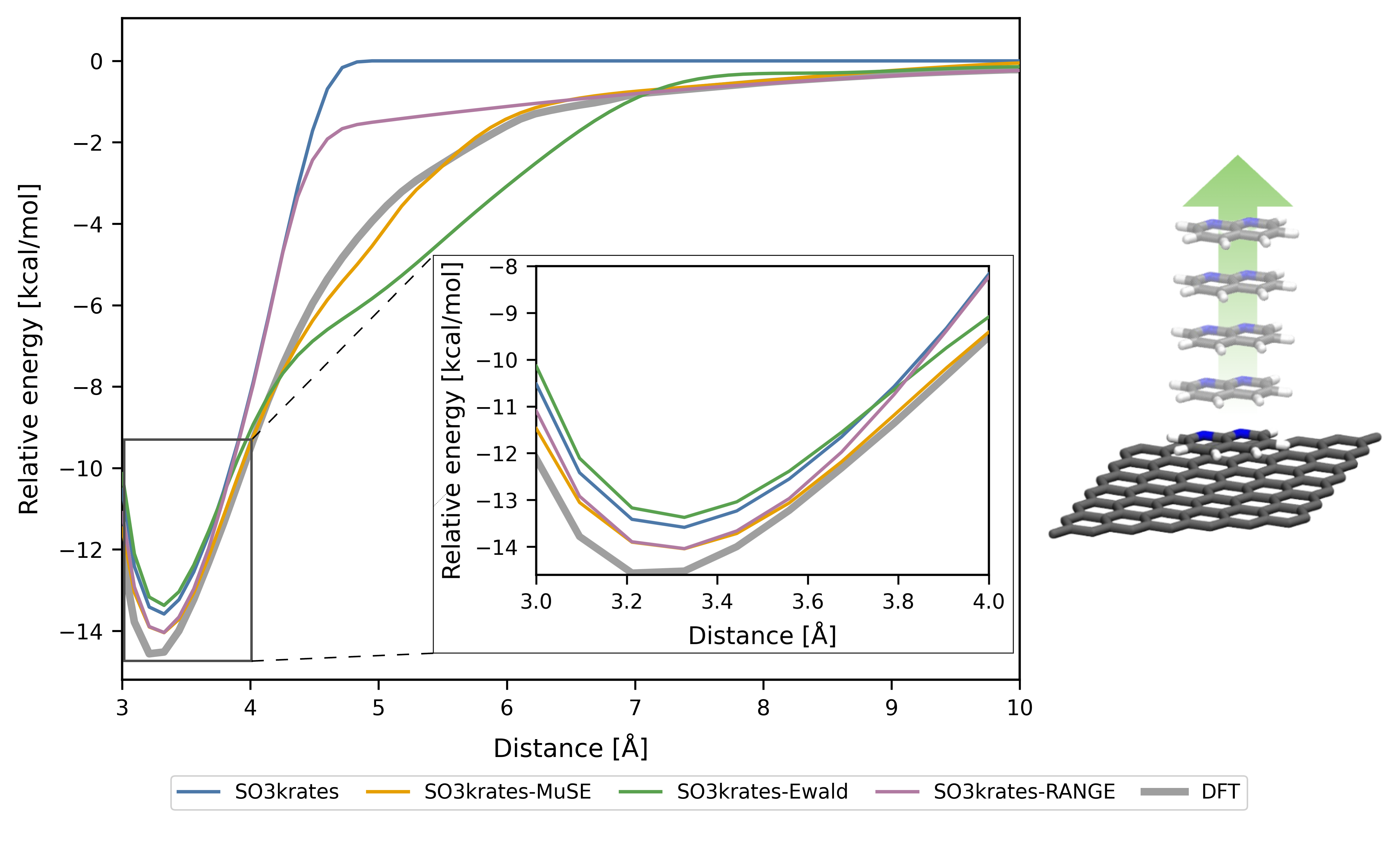}
    \caption{Dissociation curve for the 1,8-naphthyridine/graphene complex. Relative energies are reported as a function of the molecule--surface separation for SO3krates, SO3krates-MuSE, and the DFT reference.}
    \label{fig:naphthyridine_graphene_dissociation}
\end{figure}

To further probe the long-range interactions governing the dissociation of 1,8-naphthyridine from graphene, we compared the learned models against the DFT reference after training on the original TEA2023 configurations together with 1220 additional DFT structures sampled at non-equilibrium molecule--surface separations (see Methods). As shown in Fig.~\ref{fig:naphthyridine_graphene_dissociation}, the semi-local and long-range models recover the adsorption energy near the equilibrium distance relative to the dissociation limit. However, the single-scale SO3krates model systematically underestimates the depth of the adsorption well at intermediate separations, yielding a profile that is too shallow compared with PBE~\cite{pbe}+MBD-NL~\cite{mbd-nl}. This is consistent with its local 5~\AA{} radius graph, which cannot directly resolve molecule--surface couplings when relevant adsorbate and substrate atoms are separated beyond the cutoff. SO3krates-MuSE follows the DFT curve more closely across both the adsorption well and the dissociation tail, as quantified in Table~\ref{tab:dissociation_area_ranges}. RANGE also places the minimum near the DFT minimum, indicating that it captures aspects of the short-distance adsorption physics. At larger separations, however, the RANGE profile saturates and remains too attractive. This behaviour is consistent with a global communication channel that transmits non-local information, but compresses the molecule--surface coupling into a latent representation that behaves as a mean-field bottleneck. The Ewald-based model shows a different failure mode. It preserves a long-range attractive tail, but the interaction decays too slowly and remains too deep over a broad range of distances. Reciprocal-space communication therefore provides an effective global channel, but does not by itself ensure the correct real-space attenuation of dispersion-dominated adsorption interactions. MuSE avoids both limitations by constructing coarse structural representations in real space. The adsorbate and graphene sheet can therefore communicate through progressively larger coarse domains, while the atomistic branch retains the local geometry required near the adsorption minimum. Consequently, MuSE captures both the depth of the adsorption well and the smoother long-range approach to the dissociation limit.

\begin{table}[!htb]
\centering
\caption{Area between each model dissociation curve and the DFT reference, computed as $\int |E_\mathrm{model}(r)-E_\mathrm{DFT}(r)|\,dr$ over different distance ranges. Values are reported in kcal/mol \AA. Best values are shown in bold and second-best values are underlined.}
\label{tab:dissociation_area_ranges}
\begin{tabular}{lccc}
\toprule
\textbf{Method} & \textbf{Well 3--4 \AA} & \textbf{Tail 4--10 \AA} & \textbf{Total 3--10 \AA} \\
\midrule
SO3krates       & 0.9654 & 8.6498 & 9.6151 \\
SO3krates-MuSE  & \textbf{0.3481} & \textbf{0.9788} & \textbf{1.3269} \\
SO3krates-Ewald & 1.0276 & \underline{3.9690} & 4.9966 \\
SO3krates-RANGE & \underline{0.6121} & 4.3213 & \underline{4.9334} \\
\bottomrule
\end{tabular}
\end{table}

\subsection*{MuSE is model agnostic}
To assess whether the benefits of MuSE depend on the particular message-passing backbone, we evaluate three representative architectures, SO3krates~\cite{so3krates2}, MACE~\cite{mace}, and PaiNN~\cite{schutt2021painn}, on the MD22 double-walled nanotube task. For each backbone, we compare the original model with its MuSE-augmented counterpart under the same training protocol and report mean and standard deviation over four runs. Table~\ref{tab:models-mae} summarises the resulting energy and force mean absolute errors.

\begin{table}[!htb]
\centering
\caption{Energy (\textit{E}) and force (\textit{F}) MAE reported as mean $\pm$ std across four runs on the MD22 double-walled nanotube subset using different message-passing architectures. Lower is better. The best value for each backbone and metric is shown in \textbf{bold}. MuSE rows report the relative change with respect to the corresponding non-coarse baseline in parentheses; positive indicates improvement.}
\label{tab:models-mae}
\small
\begin{tabular}{lcc}
\toprule
& \multicolumn{2}{c}{\textbf{Double-walled nanotube (800)}} \\
\cmidrule(lr){2-3}
 & \textit{E} & \textit{F} \\
\textbf{Model} & $(\mathrm{kcal/mol})$ & $(\mathrm{kcal/mol/}$\AA$)$ \\
\midrule
SO3krates      & 0.986 $\pm$ 0.094 & 0.590 $\pm$ 0.010 \\
SO3krates-MuSE & \textbf{0.542 $\pm$ 0.011} (+38.0\%) & \textbf{0.434 $\pm$ 0.017} (+26.4\%) \\
\midrule
PaiNN          & 0.564 $\pm$ 0.033 & 0.471 $\pm$ 0.014 \\
PaiNN-MuSE     & \textbf{0.506 $\pm$ 0.027} (+10.2\%) & \textbf{0.370 $\pm$ 0.010} (+21.5\%) \\
\midrule
MACE           & 1.295 $\pm$ 0.077 & 0.668 $\pm$ 0.014 \\
MACE-MuSE      & \textbf{0.727 $\pm$ 0.112} (+43.9\%) & \textbf{0.432 $\pm$ 0.017} (+35.4\%) \\
\bottomrule
\end{tabular}
\end{table}

Across all three backbones, introducing the multiscale pathway yields consistent gains in both energy and force accuracy. Energy MAE improves by 38.0\% for SO3krates, 10.2\% for PaiNN, and 43.9\% for MACE, while force MAE improves by 26.4\%, 21.5\%, and 35.4\%, respectively. The fact that these improvements are observed for architecturally distinct equivariant models indicates that the benefits of MuSE do not depend on a specific message-passing design.

These results support two main conclusions. First, the advantage of MuSE is robust across different backbone families: despite substantial differences in the underlying interaction blocks, the addition of multiscale context systematically reduces both energy and force errors. Secondly, the magnitude of the improvement depends on the strength of the original single-scale model, with larger relative gains observed when the baseline backbone struggles more on this extended structure, most notably for MACE and SO3krates. This pattern suggests that the coarse-to-fine hierarchy provides complementary non-local information that is not consistently recovered by single-scale neighbourhood aggregation alone, and that MuSE therefore acts as a general enhancement mechanism rather than a model-specific modification.

\subsection*{Final remarks}

The results prove that MuSE improves the modelling of medium- and long-range interactions in MLFFs while preserving atomistic-resolution predictions. The model achieves this by constructing coarser molecular representations, processing them with the same MLFF architectures used at the atomic level, and then projecting the information back onto the atoms. This approach enables atoms to receive information from more distant regions of the system without relying solely on a larger cutoff, a deeper network, or a separate analytic long-range correction. Consequently, systematic improvements in energy and force accuracy are observed for both molecular and periodic systems, with the most significant gains occurring when non-local interactions are particularly relevant.

A comparison with alternative strategies clarifies the advantages of the multiscale construction. Increasing the cutoff allows atoms to access more neighbours, but also results in a denser graph and higher memory consumption with no substantial gains. Adding more message-passing layers increases the number of local updates, yet this approach is computationally expensive and does not necessarily enhance the recovery of long-range couplings. Explicit long-range modules can facilitate non-local communication, but their compressed or fixed forms may be unable to reproduce the distance dependence observed in the reference Hessian. In contrast, MuSE offers a comprehensive solution by shortening the communication path between distant atoms through coarse representations while maintaining an atomistic pathway for local chemical detail. The observed improvements with SO3krates, PaiNN, and MACE further suggest that this mechanism is not limited to a single model architecture.

Several limitations remain. The hierarchical structure has a finite number of levels, which may hinder the accurate description of interactions that extend beyond the range of coarser graphs. The number of scales, pooling ratios, and cutoff progression are fixed hyperparameters, and their optimal values are likely system-dependent. Future research should investigate adaptive hierarchies, scale-selection mechanisms informed by molecular geometry, and integration with long-range physical terms when interactions extend beyond the range efficiently captured by the learned hierarchy.

\section*{Methods}

\subsection*{Multiscale Structural Ensemble (MuSE)}

We propose Multiscale Structural Ensemble (MuSE), an architecture that combines representations computed at multiple scales to construct atomic embeddings for machine-learned force fields. MuSE maintains parallel scale-specific pathways and fuses them through a learned, atom-wise gating mechanism, enabling the model to balance local detail with broader structural context when predicting energies and forces. Figure~\ref{fig:architecture}a outlines the full MuSE pipeline. Starting from the atomistic graph, we construct a hierarchy of progressively coarser graph representations using Soft Coarse-Graining Pooling (SCGP). At each scale $s$, $L$ message-passing layers update the node embeddings $\tilde{\mathbf{h}}^{(s)}_i$ on the corresponding graph. The resulting coarse embeddings are then upsampled to the atomistic resolution using the SCGP weights as interpolation stencils, yielding per-atom invariant multiscale representations $\mathbf{z}^{(s)}_i$. Finally, the decoder fuses the multiscale representations into a single per-atom latent $\mathbf{f}_i$, which is then mapped to a scalar atomic energy contribution by an energy head. Figure~\ref{fig:architecture}b shows the associated increase in cut-off radius across successive scales.

To formalise the multiscale hierarchy introduced above, let $N$ denote the number of atoms in the system. For each hierarchy scale $s$, we consider a graph with $N^{(s)}$ nodes, node coordinates $\mathbf{r}_i^{(s)}\in\mathbb{R}^{3}$, the associated node features $\mathbf{h}_i^{(s)}\in\mathbb{R}^{F}$, and a cutoff radius $r_{c}^{(s)}$ to compute the neighbourhood for each node. In particular, at the atomistic level $s=0$, we have $N^{(0)} = N$; $\mathbf{r}_i^{(0)}\in\mathbb{R}^{3}$ denotes the atomic coordinates and the initial embedding $\mathbf{h}_i^{(0)}\in\mathbb{R}^{ F}$, which are obtained from an embedding for each atom type. For each scale, the cutoff radius is increased following $r_{c}^{(s+1)} = \gamma \cdot r_{c}^{(s)}$ where $\gamma > 1$, and the number of nodes are decreased following $N^{(s+1)} = \lfloor\varphi \cdot N^{(s)}+0.5\rfloor$ where $\varphi < 1$.

At each of the $s$ hierarchical scales, the current node embeddings serve as inputs to the message-passing block, which refines them into updated representations that are then used to construct the next hierarchical level via SCGP. At level $s$, the node features are updated by $L$ message-passing layers.
\begin{equation}
\mathbf{\tilde{h}}^{(s)}_{i}
=\mathcal{T}_L^{(s)}\!\bigl(\mathbf{r}_i^{(s)},\, \mathbf{h}^{(s)}_{i}\bigr),
\end{equation}
where $\mathcal{T}_L^{(s)}$ denotes the composition of the $L$ message-passing layers used at that level, and $\mathbf{\tilde{h}}^{(s)}_{i}$ are the corresponding output embedding of the node $i$.

Given the coordinates $\mathbf{r}_i^{(s)}$ and the scale $s$ feature outputs $\mathbf{\tilde{h}}_i^{(s)}$, SCGP yields a coarsened molecule with $N^{(s+1)}$ cluster nodes, a new set of coordinates $\mathbf{r}_i^{(s+1)}$ and feature node embeddings $\mathbf{h}_i^{(s+1)}$. These coarse-grained features are then used as the input for the message-passing layers at level $s+1$. The process is repeated until the $S$ scales are computed. 

Finally, each of the scales is upsampled back to the atomistic $s = 0$ scale, and the resulting S representations are combined in the decoder to produce the final $h_i$ representation for the $N$ atoms from the molecule. Then these final representations are used to predict the atomic scalar energy contribution, which, by adding up over all atoms, will give the energy of the system.

\subsection*{Soft Coarse-Graining Pooling (SCGP)}
\label{sec:coarse_graining}

We introduce Soft Coarse-Graining Pooling (SCGP), a grid-free and differentiable pooling operator for constructing coarse-grained representations. SCGP assigns atoms to cluster centres through smooth, distance-based weights that form a partition of unity, and aggregates atomic features using these weights. By replacing hard assignments with continuous weighting, SCGP yields a pooling map that varies smoothly with atomic coordinates while preserving translation invariance and $\mathrm{SO}(3)$ equivariance.

Formally, for the $j \in \{1,\dots,N^{(s)}\}$ node coordinates $\mathbf{r}_j^{(s)}$ with the respective output feature embeddings $\mathbf{\tilde{h}}_j^{(s)}$:
\begin{enumerate}
    \item Select the initial $N^{(s+1)}$ cluster positions $\mathbf{r}_i^{(s+1)}$ from the lower-scale positions $\mathbf{r}_i^{(s)}$ via Farthest Point Sampling (FPS). When periodic boundary conditions are present, FPS uses minimum-image distances (PBC-aware), i.e., the farthest-point criterion is evaluated using the shortest separation under periodic wrapping. This yields a homogeneous sampling of the full system for both periodic and non-periodic structures.
    \item For each of the $i \in \{1,\dots,N^{(s+1)}\}$ clusters, we identify the subset of nodes $k \in \mathcal{N}_i$, where $\mathcal{N}_i$ are the $\mathbf{r}_j^{(s)}$ neigbours within a radius $r_{\mathrm{pool}}$ of the initial cluster position $\mathbf{r}_i^{(s+1)}$. We then compute weights for these neighbours as the product of an exponential decay and a smooth cutoff function of the distance $d_{ik} = \|\mathbf{r}_i^{(s+1)} - \mathbf{r}_{k}^{(s)}\|$,
    \begin{equation}
        \tilde{w}^{(s)}_{ik} = e^{-d_{ik}} \cdot f_{\mathrm{cut}}(d_{ik}) \quad \text{for } k \in \mathcal{N}_i,
    \end{equation}
    and normalise them to satisfy a partition of unity over the local anchors:
    \begin{equation}
        w^{(s)}_{ik} = \frac{\tilde{w}^{(s)}_{ik}}{\sum_{k \in \mathcal{N}_i} \tilde{w}^{(s)}_{ik'} }.
    \end{equation}
    \item Refine the positions $\mathbf{r}_i^{(s+1)}$ and create feature vector $\mathbf{h}_i^{(s+1)}$ of each $N^{(s+1)}$ cluster by weighted aggregation over its assigned $k \in \mathcal{N}_i$ subset. Then:
    \begin{equation}
    \mathbf{r}^{(s+1)}_i
    = \sum\limits_{k \in \mathcal{N}_i} w^{(s)}_{ik}\,\mathbf{r}_k^{(s)},
    \qquad
    \mathbf{h}^{(s+1)}_i
    = \sum\limits_{k \in \mathcal{N}_i} w^{(s)}_{ik}\,\mathbf{\tilde{h}}_k^{(s)}.
    \label{eq:aggregation}
    \end{equation}
\end{enumerate}

\begin{figure}[!htb]
    \centering
    \includegraphics[width=\linewidth]{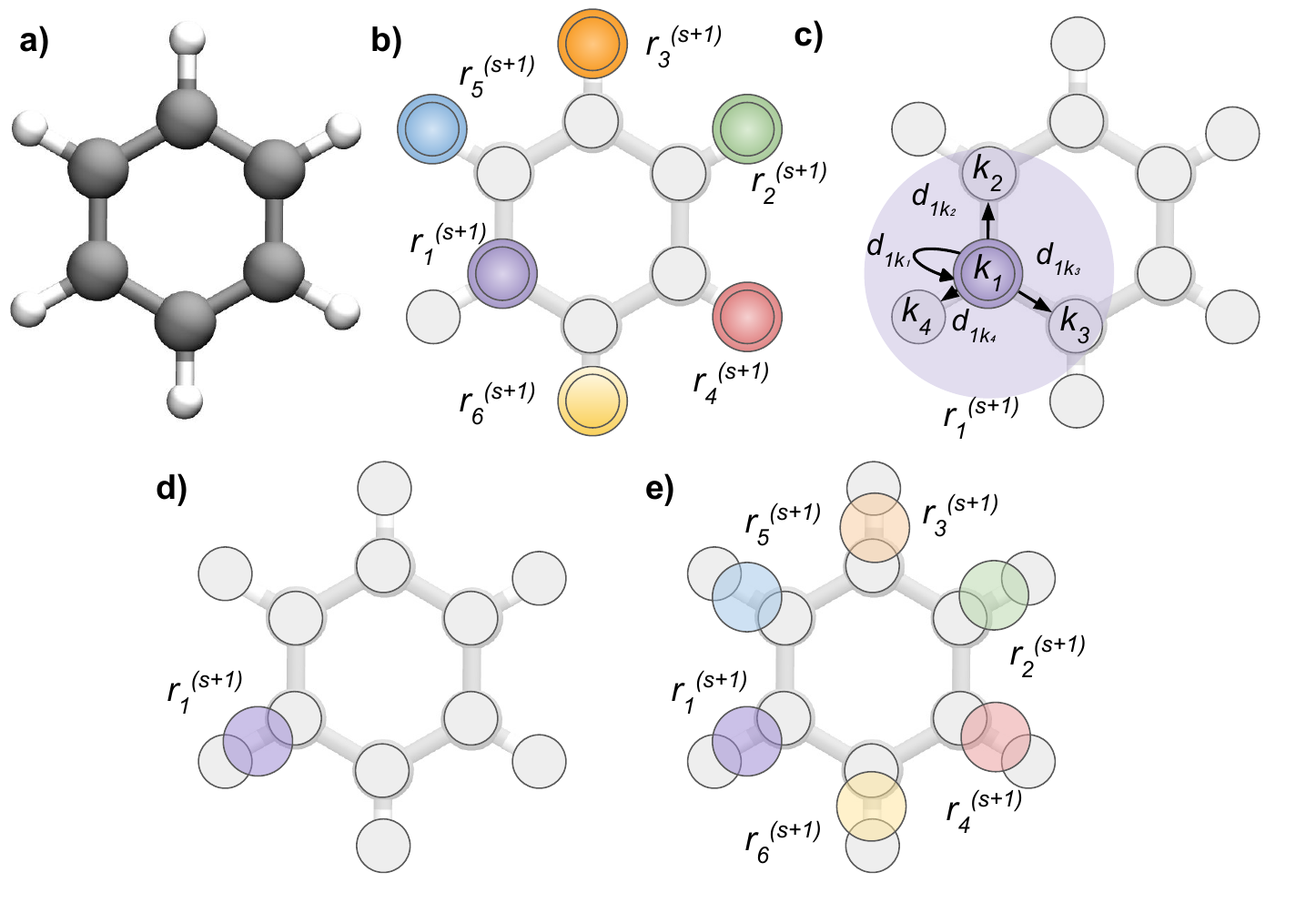}
    \caption{Soft Coarse-Graining Pooling (SCGP). \textbf{a)} Benzene molecule used for illustration. \textbf{b)} Initial cluster positions $\mathbf{r}^{(s+1)}_i$ are selected from the lower-scale coordinates $\mathbf{r}^{(s)}_j$ via Farthest Point Sampling (FPS). \textbf{c)} For a given cluster $i$ (yellow), neighbours $k \in \mathcal{N}_i$ are identified within the pooling radius $r_{\mathrm{pool}}$ (green region), and unnormalised weights $\tilde{w}^{(s)}_{ik}$ are computed from distances $d_{ik}$. \textbf{d)} Illustration of the resulting local soft assignment for cluster $i$, where nearby nodes receive higher weight. \textbf{e)} Repeating the weighting and aggregation for each initial $\mathbf{r}^{(s+1)}_i$ yields the refined coarse positions $\mathbf{r}^{(s+1)}_i$.}
    \label{fig:NFPS}
\end{figure}

This procedure defines a hierarchy of progressively coarser graphs. In particular, the weighted aggregation in Eq.\ref{eq:aggregation} maps the scale-$s$ representation $\mathbf{r}^{(s)}_j, \mathbf{\tilde{h}}^{(s)}_j$ to a new set of $N^{(s+1)}$ coarse nodes at scale $s+1$, with positions $\mathbf{r}^{(s+1)}_i$ and features $\mathbf{h}^{(s+1)}_i$. Each coarse node $i$ can be interpreted as a soft cluster of its assigned fine-scale neighbours $k \in \mathcal{N}_i$: the position $\mathbf{r}^{(s+1)}_i$ is the weighted barycentre of the corresponding fine coordinates, and the feature vector $\mathbf{h}^{(s+1)}_i$ is the associated weighted summary of the fine embeddings. Consequently, the scale $s+1$ graph provides a higher-level description of the same structure, in which groups of scale $s$ nodes are (softly) merged into fewer representative nodes while preserving smooth dependence on geometry. The weights $w^{(s)}_{ik}$ are stored and reused in the decoder to upsample coarse-scale embeddings back to finer resolutions. Figure~\ref{fig:NFPS} illustrates the SCGP algorithm: a) the initial $\mathbf{r}_i^{(s+1)}$ coordinates based on FPS sampling, b) how the weights for the local soft assignment are computed, and c) the resulting refined coarse-node coordinates based on the weighting. During training, we randomise the initial FPS seed to expose the model to diverse coarsening geometries. During molecular dynamics, the initial FPS positions are fixed, while the soft assignments vary smoothly with the evolving geometry.

\subsection*{Upsampling}
\label{subsec:upsampling}

To fuse multiscale information at the atomistic resolution, we upsample embeddings computed at coarser scales back to finer scales by reusing the SCGP pooling weights $w^{(s)}_{ik}$ stored during coarse-graining (see SCGP). For each scale transition $(s \rightarrow s+1)$, coarse-graining associates every coarse node $k \in \{1,\dots,N^{(s+1)}\}$ with a subset of finer-scale nodes $i \in  \{1,\dots,N^{(s)}\}$ and corresponding weights $w^{(s)}_{ik}$. For convenience, we also introduce the reverse (fine-to-coarse) neighbourhood,
\begin{equation}
\mathcal{N}^{-1}_i = \{\, k \in \{1,\dots,N^{(s+1)}\} \;:\; i \in \mathcal{N}_k \,\},
\label{eq:reverse_neighbourhood}
\end{equation}
which collects all coarse nodes whose pooling neighbourhood contains the fine node $i$.

In the decoder, we reuse the coefficients $w^{(s)}_{k i}$ as an interpolation stencil and define a pairwise upsampling operator $\mathcal{U}^{(s+1 \rightarrow s)}$ that distributes coarse-scale embeddings to the finer scale by weighted accumulation over $\mathcal{N}^{-1}_i$. Formally, $\mathcal{U}^{(s+1 \rightarrow s)}$ is defined element-wise by
\begin{equation}
\mathcal{U}^{(s+1 \rightarrow s)}(\mathbf{\tilde{h}}^{(s+1)})_i
=
\sum_{k \in \mathcal{N}^{-1}_i}
w^{(s)}_{ik}\,\mathbf{\tilde{h}}^{(s+1)}_k,
\qquad
i = 1,\dots,N^{(s)}.
\label{eq:upsampling_pairwise}
\end{equation}
This operation corresponds to a weighted scatter from the coarse nodes to the finer nodes, reusing the same coefficients $w^{(s)}_{ik}$ that were employed during coarse-graining.

In the decoder, we require an atom-wise representation for each scale. Accordingly, we define the multi-step upsampling operator from scale $s$ to the atomistic resolution $0$ as the composition of the pairwise operators (where $\circ$ denotes operator composition):
\begin{equation}
\mathcal{U}^{(s \rightarrow 0)} = (
\mathcal{U}^{(1\rightarrow 0)} \circ \mathcal{U}^{(2\rightarrow 1)} \circ \cdots \circ \mathcal{U}^{(s\rightarrow s-1)} )(\mathbf{\tilde{h}}_k^{(s)}),
\qquad s = 1,\dots,S-1,
\label{eq:upsampling_operator_def}
\end{equation}
and set
\begin{equation}
\begin{gathered}
\mathbf{z}_i^{(0)} = \tilde{\mathbf{h}}_i^{(0)}, \qquad
\mathbf{z}_i^{(s)} = \mathcal{U}^{(s\rightarrow 0)}\bigl(\tilde{\mathbf{h}}_k^{(s)}\bigr), \\
s = 1,\dots,S-1,\quad i = 1,\dots,N^{(0)},\quad k = 1,\dots,N^{(s)}.
\end{gathered}
\label{eq:upsampling_to_atomistic}
\end{equation}
By construction, $\mathbf{z}_i^{(s)}$ is defined on the same set of $N^{(0)}$ atoms for all scales and can be fused in the subsequent decoding stage.

\subsection*{Decoding}
\label{subsec:decoding}

Given the multiscale atom-wise embeddings $\mathbf{z}_i^{(s)}$  for each atom $i \in \{1,\dots,N^{(0)}\}$, the decoder fuses them through an atom-wise gated mixture across scales.

First, we concatenate the representations across the $S$ scales:
\begin{equation}
\mathbf{c_i} =
\bigl[\mathbf{z}_i^{(0)} \,\Vert\, \mathbf{z}_i^{(1)} \,\Vert\, \cdots \,\Vert\, \mathbf{z}_i^{(s)} \bigr]
\in \mathbb{R}^{SF}.
\label{eq:decoder_concat}
\end{equation}
A linear map produces scale logits $\mathbf{g}_i \in \mathbb{R}^{S}$:
\begin{equation}
\mathbf{g}_i = \mathbf{W}\,\mathbf{c}_i + \mathbf{b},
\qquad
\mathbf{W} \in \mathbb{R}^{S \times S F},\;\; \mathbf{b} \in \mathbb{R}^{S}.
\label{eq:decoder_logits}
\end{equation}
The corresponding mixture coefficients $\alpha_{i,s}$ are obtained by softmax normalisation:
\begin{equation}
\alpha_{i,s} =
\frac{\exp(g_{i,s})}{\sum_{s'=0}^{S-1} \exp(g_{i,s'})},
\qquad s = 0,\dots,S-1.
\label{eq:decoder_softmax}
\end{equation}

Finally, the fused atom-wise representation is the weighted combination
\begin{equation}
\mathbf{f}_i =  \phi\!\left(\sum_{s=0}^{S-1} \alpha_{i,s}\, \mathbf{z}_i^{(s)}\right),
\label{eq:decoder_fusion}
\end{equation}
where $\phi(\cdot)$ denotes an element-wise activation function.
The coefficients $\alpha_{i,s}$ quantify the contribution of each scale to atom $i$ in the final latent representation, providing an interpretable routing of information across the hierarchy.

\subsection*{Energy and Force prediction}
\label{subsec:energy_prediction}

The decoder produces a final atom-wise latent representation $\mathbf{f}_i$ for each atom $i \in \{1,\dots,N^{(0)}\}$. To predict the total potential energy, we employ a  energy head $E_{\mathrm{head}}$, implemented as a multilayer perceptron (MLP), which maps each atom embedding to a scalar energy contribution:

\begin{equation}
e_i = E_{\mathrm{head}}(\mathbf{f}_i),
\qquad
e_i \in \mathbb{R}^1.
\label{eq:atomic_energy}
\end{equation}

The predicted system energy is then obtained by an atom-wise sum,

\begin{equation}
\hat{E} = \sum_{i=1}^{N} e_i,
\label{eq:total_energy}
\end{equation}

To enable force prediction while ensuring that the forces are conservative, we obtain the predicted forces as the negative gradient of the predicted total energy with respect to the atomic coordinates:
\begin{equation}
\hat{\mathbf{F}}_i = -\,\frac{\partial \hat{E}}{\partial \mathbf{r}_i},
\qquad i = 1,\dots,N.
\label{eq:forces_from_energy}
\end{equation}

\subsection*{Training and MD simulations}

All models were trained with joint energy and force supervision using dataset-specific protocols; full training hyperparameters, data splits, optimisation schedules, and additional simulation details are provided in the Supplementary Information, Section 1 (Experimental Details). For Fig.~\ref{fig:vdos-alanine}a, the canonical $\alpha$-helical alanine structure was relaxed with L-BFGS~\cite{lbfgs} until the force norm fell below $10^{-4}$, after which velocities were sampled from a Maxwell--Boltzmann distribution at 600~K and a 1~ns NVE trajectory was propagated with Velocity-Verlet using a 0.2~fs time step. The unfolded Ace-Ala$_{15}$-NMe trajectory in Fig.~\ref{fig:vdos-alanine}b,c,d was prepared with the same relaxation criterion and velocity initialisation and then run for 500~ps with a 0.5~fs time step using a Hoover thermostat at 300~K.

For the adsorbate–graphene model, all calculations were performed with FHI-aims (250320)~\cite{blum2009fhi} using the PBE~\cite{pbe} exchange-correlation functional with the non-local many-body dispersion correction (MBD-NL)~\cite{mbd-nl}, the "light" 2020 species defaults. The equilibrium structure was obtained by full BFGS relaxation to a maximum residual force of 5 × 10\textsuperscript{-3} eV/Å. From this reference, a dissociation curve was sampled at 148 surface-normal separations between 0.9 equilibrium distance and 20 Å equispaced points, with the graphene frozen and adsorbate atoms constrained to a fixed-z plane (free in xy). To probe non-equilibrium configurations, an additional 1220-structure dataset was generated on a 0.25 Å grid from 5.0 to 20.0 Å (61 distances) with 20 uniformly-sampled SO(3) orientations per distance; each structure was relaxed with only the adsorbate z-coordinates fixed, allowing both graphene and in-plane molecular degrees of freedom to relax. TEA2023 conformations were recomputed with the same parameters to ensure consistency. Control files and training data are provided in the supplementary.

\section*{Code Availability}
The code for MuSE and the associated experiments will be released on GitHub upon acceptance.

\section*{Data Availability}
The datasets used in this work are publicly available as follows: GEMS alanine-peptide dataset~\cite{GEMS}, MD22 dataset~\cite{MD22}, and TEA2023 dataset~\cite{crashtest}.
The DFT PBE0+MBD hessians from the alanine peptide can be found at \cite{hessians_sergio}.
The hessian files, dissociation curves, and the MD trajectories used for this work will be published on Zenodo upon acceptance.

\bibliography{refs}

@inproceedings{ewald_sum,
  title={Ewald-based long-range message passing for molecular graphs},
  author={Kosmala, Arthur and Gasteiger, Johannes and Gao, Nicholas and G{\"u}nnemann, Stephan},
  booktitle={International Conference on Machine Learning},
  pages={17544--17563},
  year={2023},
  organization={PMLR}
}

@article{king2025chargeslongrange,
  title     = {Machine learning of charges and long-range interactions from energies and forces},
  author    = {King, Daniel S. and Kim, Dongjin and Zhong, Peichen and Cheng, Bingqing},
  journal   = {Nature Communications},
  volume    = {16},
  pages     = {8763},
  year      = {2025},
}

@inproceedings{zhou2023unimol,
  title     = {{Uni-Mol}: A Universal {3D} Molecular Representation Learning Framework},
  author    = {Zhou, Gengmo and Gao, Zhifeng and Ding, Qiankun and Zheng, Hang and Xu, Hongteng and Wei, Zhewei and Zhang, Linfeng and Ke, Guolin},
  booktitle = {International Conference on Learning Representations},
  year      = {2023},
}

@article{mendezlucio2024mole,
  title     = {{MolE}: a foundation model for molecular graphs using disentangled attention},
  author    = {M{\'e}ndez-Lucio, Oscar and Nicolaou, Christos A. and Earnshaw, Berton},
  journal   = {Nature Communications},
  volume    = {15},
  pages     = {9431},
  year      = {2024},
}

@inproceedings{gasteiger_dimenet_2020,
  title = {Directional Message Passing for Molecular Graphs},
  author = {Gasteiger, Johannes and Gro{\ss}, Janek and G{\"u}nnemann, Stephan},
  booktitle={International Conference on Learning Representations (ICLR)},
  year = {2020}
}

@inproceedings{gilmer2017neural,
  title     = {Neural Message Passing for Quantum Chemistry},
  author    = {Gilmer, Justin and Schoenholz, Samuel S. and Riley, Patrick F. and Vinyals, Oriol and Dahl, George E.},
  booktitle = {Proceedings of the 34th International Conference on Machine Learning},
  series    = {Proceedings of Machine Learning Research},
  volume    = {70},
  pages     = {1263--1272},
  year      = {2017},
  publisher = {PMLR},
  address   = {Sydney, Australia}
}

@inproceedings{gasteiger_dimenetpp_2020,
title = {Fast and Uncertainty-Aware Directional Message Passing for Non-Equilibrium Molecules},
author = {Gasteiger, Johannes and Giri, Shankari and Margraf, Johannes T. and G{\"u}nnemann, Stephan},
booktitle={Machine Learning for Molecules Workshop, NeurIPS},
year = {2020} }

@article{qi2017pointnet++,
  title={{PointNet++}: Deep Hierarchical Feature Learning on Point Sets in a Metric Space},
  author={Qi, Charles Ruizhongtai and Yi, Li and Su, Hao and Guibas, Leonidas J},
  journal={Advances in Neural Information Processing Systems},
  volume={30},
  year={2017}
}

@inproceedings{qi2017pointnet,
  title={{PointNet}: Deep Learning on Point Sets for {3D} Classification and Segmentation},
  author={Qi, Charles R and Su, Hao and Mo, Kaichun and Guibas, Leonidas J},
  booktitle={Proceedings of the IEEE Conference on Computer Vision and Pattern Recognition},
  pages={652--660},
  year={2017}
}

@article{wang2019ddgcn,
  title={Dynamic graph cnn for learning on point clouds},
  author={Wang, Yue and Sun, Yongbin and Liu, Ziwei and Sarma, Sanjay E and Bronstein, Michael M and Solomon, Justin M},
  journal={ACM Transactions on Graphics (tog)},
  volume={38},
  number={5},
  pages={1--12},
  year={2019},
  publisher={Acm New York, NY, USA}
}

@INPROCEEDINGS{kpconv,
  author={Thomas, Hugues and Qi, Charles R. and Deschaud, Jean-Emmanuel and Marcotegui, Beatriz and Goulette, François and Guibas, Leonidas},
  booktitle={2019 IEEE/CVF International Conference on Computer Vision (ICCV)}, 
  title={{KPConv}: Flexible and Deformable Convolution for Point Clouds},
  year={2019},
  volume={},
  number={},
  pages={6410-6419},
}

@inproceedings{hu2020randlanet,
  title     = {{RandLA-Net}: Efficient Semantic Segmentation of Large-Scale Point Clouds},
  author    = {Hu, Qingyong and Yang, Bo and Xie, Linhai and Rosa, Stefano and Guo, Yulan and Wang, Zhihua and Trigoni, Niki and Markham, Andrew},
  booktitle = {Proceedings of the IEEE/CVF Conference on Computer Vision and Pattern Recognition (CVPR)},
  year      = {2020},
}

@inproceedings{zhao2021pointtransformer,
    author    = {Zhao, Hengshuang and Jiang, Li and Jia, Jiaya and Torr, Philip H.S. and Koltun, Vladlen},
    title     = {Point Transformer},
    booktitle = {Proceedings of the IEEE/CVF International Conference on Computer Vision (ICCV)},
    month     = {October},
    year      = {2021},
    pages     = {16259-16268}
}

@inproceedings{landrieu2018spg,
  author={Landrieu, Loic and Simonovsky, Martin},
  booktitle={2018 IEEE/CVF Conference on Computer Vision and Pattern Recognition}, 
  title={Large-Scale Point Cloud Semantic Segmentation with Superpoint Graphs}, 
  year={2018},
  volume={},
  number={},
  pages={4558-4567},
}

@article{ewald_sum_2,
  title={Latent {Ewald} summation for machine learning of long-range interactions},
  author={Cheng, Bingqing},
  journal={npj Computational Materials},
  volume={11},
  number={1},
  pages={80},
  year={2025},
  publisher={Nature Publishing Group UK London}
}

@article{nequip,
  title={E (3)-equivariant graph neural networks for data-efficient and accurate interatomic potentials},
  author={Batzner, Simon and Musaelian, Albert and Sun, Lixin and Geiger, Mario and Mailoa, Jonathan P and Kornbluth, Mordechai and Molinari, Nicola and Smidt, Tess E and Kozinsky, Boris},
  journal={Nature Communications},
  volume={13},
  number={1},
  pages={2453},
  year={2022},
  publisher={Nature Publishing Group UK London}
}

@article{allegro,
  title={Learning local equivariant representations for large-scale atomistic dynamics},
  author={Musaelian, Albert and Batzner, Simon and Johansson, Anders and Sun, Lixin and Owen, Cameron J and Kornbluth, Mordechai and Kozinsky, Boris},
  journal={Nature Communications},
  volume={14},
  number={1},
  pages={579},
  year={2023},
  publisher={Nature Publishing Group UK London}
}

@article{alignn,
  title={Atomistic line graph neural network for improved materials property predictions},
  author={Choudhary, Kamal and DeCost, Brian},
  journal={npj Computational Materials},
  volume={7},
  number={1},
  pages={185},
  year={2021},
  publisher={Nature Publishing Group UK London}
}

@article{ko2021fourth,
  title={A fourth-generation high-dimensional neural network potential with accurate electrostatics including non-local charge transfer},
  author={Ko, Tsz Wai and Finkler, Jonas A and Goedecker, Stefan and Behler, J{\"o}rg},
  journal={Nature Communications},
  volume={12},
  number={1},
  pages={398},
  year={2021},
  publisher={Nature Publishing Group UK London}
}

@article{gasteiger2021gemnet,
  title={{GemNet}: Universal Directional Graph Neural Networks for Molecules},
  author={Gasteiger, Johannes and Becker, Florian and G{\"u}nnemann, Stephan},
  journal={Advances in Neural Information Processing Systems},
  volume={34},
  pages={6790--6802},
  year={2021}
}

@article{kabylda2023efficient,
  title={Efficient interatomic descriptors for accurate machine learning force fields of extended molecules},
  author={Kabylda, Adil and Vassilev-Galindo, Valentin and Chmiela, Stefan and Poltavsky, Igor and Tkatchenko, Alexandre},
  journal={Nature Communications},
  volume={14},
  number={1},
  pages={3562},
  year={2023},
  publisher={Nature Publishing Group UK London}
}

@article{unke2021spookynet,
  title={{SpookyNet}: Learning force fields with electronic degrees of freedom and nonlocal effects},
  author={Unke, Oliver T and Chmiela, Stefan and Gastegger, Michael and Sch{\"u}tt, Kristof T and Sauceda, Huziel E and M{\"u}ller, Klaus-Robert},
  journal={Nature Communications},
  volume={12},
  number={1},
  pages={7273},
  year={2021},
  publisher={Nature Publishing Group UK London}
}

@inproceedings{conformer,
  title={Complete and Efficient Graph Transformers for Crystal Material Property Prediction},
  author={Yan, Keqiang and Fu, Cong and Qian, Xiaofeng and Qian, Xiaoning and Ji, Shuiwang},
  booktitle={International Conference on Learning Representations},
  year={2024}
}

@article{matformer,
  title={Periodic graph transformers for crystal material property prediction},
  author={Yan, Keqiang and Liu, Yi and Lin, Yuchao and Ji, Shuiwang},
  journal={Advances in Neural Information Processing Systems},
  volume={35},
  pages={15066--15080},
  year={2022}
}

@inproceedings{potnet,
  title={Efficient approximations of complete interatomic potentials for crystal property prediction},
  author={Lin, Yuchao and Yan, Keqiang and Luo, Youzhi and Liu, Yi and Qian, Xiaoning and Ji, Shuiwang},
  booktitle={International Conference on Machine Learning},
  pages={21260--21287},
  year={2023},
  organization={PMLR}
}

@article{schnet,
  title={{SchNet}: A Continuous-Filter Convolutional Neural Network for Modeling Quantum Interactions},
  author={Sch{\"u}tt, Kristof and Kindermans, Pieter-Jan and Sauceda Felix, Huziel Enoc and Chmiela, Stefan and Tkatchenko, Alexandre and M{\"u}ller, Klaus-Robert},
  journal={Advances in Neural Information Processing Systems},
  volume={30},
  year={2017}
}

@Article{cartnet,
author ="Solé, \`Alex and Mosella-Montoro, Albert and Cardona, Joan and Gómez-Coca, Silvia and Aravena, Daniel and Ruiz, Eliseo and Ruiz-Hidalgo, Javier",
title  ="A Cartesian encoding graph neural network for crystal structure property prediction: application to thermal ellipsoid estimation",
journal  ="Digital Discovery",
year  ="2025",
volume  ="4",
issue  ="3",
pages  ="694-710",
publisher  ="RSC",
}

@inproceedings{schutt2021painn,
  title={Equivariant message passing for the prediction of tensorial properties and molecular spectra},
  author={Sch{\"u}tt, Kristof and Unke, Oliver and Gastegger, Michael},
  booktitle={International Conference on Machine Learning},
  pages={9377--9388},
  year={2021},
  organization={PMLR}
}

@inproceedings{mace,
  title={{MACE}: Higher Order Equivariant Message Passing Neural Networks for Fast and Accurate Force Fields},
  author={Batatia, Ilyes and Kovacs, David P and Simm, Gregor and Ortner, Christoph and Cs{\'a}nyi, G{\'a}bor},
  booktitle={Advances in Neural Information Processing Systems},
  volume={35},
  pages={11423--11436},
  year={2022}
}

@inproceedings{spotr,
  title={Self-positioning point-based transformer for point cloud understanding},
  author={Park, Jinyoung and Lee, Sanghyeok and Kim, Sihyeon and Xiong, Yunyang and Kim, Hyunwoo J},
  booktitle={Proceedings of the IEEE/CVF conference on computer vision and pattern recognition},
  pages={21814--21823},
  year={2023}
}

@article{frank2022so3krates,
  title={{SO3krates}: Equivariant Attention for Interactions on Arbitrary Length-Scales in Molecular Systems},
  author={Frank, Thorben and Unke, Oliver and M{\"u}ller, Klaus-Robert},
  journal={Advances in Neural Information Processing Systems},
  volume={35},
  pages={29400--29413},
  year={2022}
}

@article{so3lr,
  title   = {Molecular Simulations with a Pretrained Neural Network and Universal Pairwise Force Fields},
  author  = {Kabylda, Adil and Frank, J. Thorben and Su{\'a}rez-Dou, Sergio and Khabibrakhmanov, Almaz and Medrano Sandonas, Leonardo and Unke, Oliver T. and Chmiela, Stefan and M{\"u}ller, Klaus-Robert and Tkatchenko, Alexandre},
  journal = {Journal of the American Chemical Society},
  volume  = {147},
  number  = {37},
  pages   = {33723--33734},
  year    = {2025},
}

@article{so3krates2,
  title={A Euclidean transformer for fast and stable machine learned force fields},
  author={Frank, Thorben and Unke, Oliver and M{\"u}ller, Klaus-Robert and Chmiela, Stefan},
  journal={Nature Communications},
  volume={15},
  number={1},
  pages={6539},
  year={2024}
}

@article{mbd,
  title={Accurate and Efficient Method for Many-Body {van der Waals} Interactions},
  author={Tkatchenko, Alexandre and DiStasio Jr, Robert A and Car, Roberto and Scheffler, Matthias},
  journal={Physical Review Letters},
  volume={108},
  number={23},
  pages={236402},
  year={2012},
  publisher={APS}
}

@article{mbd-nl,
  title = {Density Functional Model for {van der Waals} Interactions: Unifying Many-Body Atomic Approaches with Nonlocal Functionals},
  author = {Hermann, Jan and Tkatchenko, Alexandre},
  journal = {Phys. Rev. Lett.},
  volume = {124},
  issue = {14},
  pages = {146401},
  numpages = {7},
  year = {2020},
  month = {Apr},
  publisher = {American Physical Society},
}

@article{pbe,
  title   = {Generalized Gradient Approximation Made Simple},
  author  = {Perdew, John P. and Burke, Kieron and Ernzerhof, Matthias},
  journal = {Physical Review Letters},
  volume  = {77},
  number  = {18},
  pages   = {3865--3868},
  year    = {1996},
}

@article{MD22,
author = {Chmiela, Stefan and Vassilev-Galindo, Valentin and Unke, Oliver T. and Kabylda, Adil and Sauceda, Huziel E. and Tkatchenko, Alexandre and M{\"u}ller, Klaus-Robert},
title = {Accurate Global Machine Learning Force Fields for Molecules with Hundreds of Atoms},
journal = {Science Advances},
volume = {9},
number = {2},
pages = {eadf0873},
year = {2023},
}

@article{pbe0,
    author = {Adamo, Carlo and Barone, Vincenzo},
    title = {Toward Reliable Density Functional Methods without Adjustable Parameters: The {PBE0} Model},
    journal = {The Journal of Chemical Physics},
    volume = {110},
    number = {13},
    pages = {6158-6170},
    year = {1999},
    month = {04},
    issn = {0021-9606},
}

@article{MOSELLAMONTORO2021,
title = {{2D--3D} Geometric Fusion Network Using Multi-Neighbourhood Graph Convolution for {RGB-D} Indoor Scene Classification},
journal = {Information Fusion},
volume = {76},
pages = {46-54},
year = {2021},
issn = {1566-2535},
author = {Albert Mosella-Montoro and Javier Ruiz-Hidalgo},
}

@ARTICLE{FPS,
  author={Eldar, Y. and Lindenbaum, M. and Porat, M. and Zeevi, Y.Y.},
  journal={IEEE Transactions on Image Processing}, 
  title={The farthest point strategy for progressive image sampling}, 
  year={1997},
  volume={6},
  number={9},
  pages={1305-1315},
}

@article{voxel,
  title   = {Voxel-Based Representation of {3D} Point Clouds: Methods, Applications, and Its Potential Use in the Construction Industry},
  author  = {Xu, Yusheng and Tong, Xiaohua and Stilla, Uwe},
  journal = {Automation in Construction},
  volume  = {126},
  pages   = {103675},
  year    = {2021},
  issn    = {0926-5805},
}

@Article{crashtest,
author ="Poltavsky, Igor and Charkin-Gorbulin, Anton and Puleva, Mirela and Fonseca, Grégory and Batatia, Ilyes and Browning, Nicholas J. and Chmiela, Stefan and Cui, Mengnan and Frank, J. Thorben and Heinen, Stefan and Huang, Bing and Käser, Silvan and Kabylda, Adil and Khan, Danish and Müller, Carolin and Price, Alastair J. A. and Riedmiller, Kai and Töpfer, Kai and Ko, Tsz Wai and Meuwly, Markus and Rupp, Matthias and Csányi, Gábor and von Lilienfeld, O. Anatole and Margraf, Johannes T. and Müller, Klaus-Robert and Tkatchenko, Alexandre",
title  ="Crash testing machine learning force fields for molecules{,} materials{,} and interfaces: model analysis in the {TEA} Challenge 2023",
journal  ="Chem. Sci.",
year  ="2025",
volume  ="16",
issue  ="8",
pages  ="3720-3737",
publisher  ="The Royal Society of Chemistry",
}

@article{GEMS,
  title   = {Biomolecular dynamics with machine-learned quantum-mechanical force fields trained on diverse chemical fragments},
  author  = {Unke, Oliver T. and St{\"o}hr, Martin and Ganscha, Stefan and Unterthiner, Thomas and Maennel, Hartmut and Kashubin, Sergii and Ahlin, Daniel and Gastegger, Michael and Medrano Sandon{\'a}s, Leonardo and Berryman, Joshua T. and Tkatchenko, Alexandre and M{\"u}ller, Klaus-Robert},
  journal = {Science Advances},
  year    = {2024},
  volume  = {10},
  number  = {14},
  pages   = {eadn4397},
}

@article{PRISM,
  author  = {Sol{\'e}, {\`A}lex and Mosella-Montoro, Albert and Cardona, Joan and Aravena, Daniel and G{\'o}mez-Coca, Silvia and Ruiz, Eliseo and Ruiz-Hidalgo, Javier},
  title   = {{PRISM}: Periodic Representation with Multiscale and Similarity Graph Modelling for Enhanced Crystal Structure Property Prediction},
  journal = {npj Computational Materials},
  year    = {2026},
  issn    = {2057-3960}
}

@article{lbfgs,
  title   = {On the limited memory {BFGS} method for large scale optimization},
  author  = {Liu, Dong C. and Nocedal, Jorge},
  journal = {Mathematical Programming},
  volume  = {45},
  pages   = {503--528},
  year    = {1989},
}

@article{range,
  title        = {Extending the range of graph neural networks with global encodings},
  author       = {Caruso, Alessandro and Venturin, Jacopo and Giambagli, Lorenzo and Rolando, Edoardo and El-Machachi, Zakariya and No{\'e}, Frank and Clementi, Cecilia},
  journal      = {Nature Communications},
  year         = {2026},
  volume       = {17},
  pages        = {1855},
}

@article{ZBLrepulsion,
  author  = {Ziegler, James F. and Ziegler, M. D. and Biersack, J. P.},
  title   = {SRIM---The Stopping and Range of Ions in Matter (2010)},
  journal = {Nuclear Instruments and Methods in Physics Research Section B: Beam Interactions with Materials and Atoms},
  volume  = {268},
  number  = {11--12},
  pages   = {1818--1823},
  year    = {2010},
}

@article{ase,
  title   = {The atomic simulation environment---a {P}ython library for working with atoms},
  author  = {Larsen, Ask Hjorth and Mortensen, Jens J{\o}rgen and Blomqvist, Jakob and Castelli, Ivano E. and Christensen, Rune and Du{\l}ak, Marcin and Friis, Jesper and Groves, Michael N. and Hammer, Bj{\o}rk and Hargus, Cory and others},
  journal = {Journal of Physics: Condensed Matter},
  volume  = {29},
  number  = {27},
  pages   = {273002},
  year    = {2017},
}

@INPROCEEDINGS{amsgrad,
  author={Tan, Tao and Yin, Shiqun and Liu, Kunling and Wan, Man},
  booktitle={2019 IEEE 31st International Conference on Tools with Artificial Intelligence (ICTAI)}, 
  title={On the Convergence Speed of AMSGRAD and Beyond}, 
  year={2019},
  volume={},
  number={},
  pages={464-470},
}

@article{fastattention_frank,
  author    = {Frank, J. Thorben and Chmiela, Stefan and M{\"u}ller, Klaus-Robert and Unke, Oliver T.},
  title     = {Machine learning global atomic representations with Euclidean fast attention},
  journal   = {Nature Machine Intelligence},
  year      = {2026},
  month     = mar,
  volume    = {8},
  number    = {3},
  pages     = {388--402},
  issn      = {2522-5839},
}

@article{capone2025stgnn,
  title   = {Spatio-temporal prediction using graph neural networks: A survey},
  author  = {Capone, Vincenzo and Casolaro, Angelo and Camastra, Francesco},
  journal = {Neurocomputing},
  volume  = {643},
  pages   = {130400},
  year    = {2025},
}

@article{sanderse2025multiscale,
  title   = {Scientific machine learning for closure models in multiscale problems: A review},
  author  = {Sanderse, Benjamin and Stinis, Panos and Maulik, Romit and Ahmed, Shady E.},
  journal = {Foundations of Data Science},
  volume  = {7},
  number  = {1},
  pages   = {1--62},
  year    = {2025},
}

@article{reiser2022gnnchem,
  title   = {Graph neural networks for materials science and chemistry},
  author  = {Reiser, Patrick and Neubert, Marlen and Eberhard, Andr{\'e} and Torresi, Luca and Zhou, Chen and Shao, Chen and Metni, Houssam and van Hoesel, Clint and Schopmans, Henrik and Sommer, Timo and Friederich, Pascal},
  journal = {Communications Materials},
  volume  = {3},
  number  = {1},
  pages   = {93},
  year    = {2022},
}

@misc{hessians_sergio,
      title={Stability and Vibrations of Proteins in Vacuum and Water: Bridging Quantum Accuracy and Force-Field Efficiency}, 
      author={Sergio Suárez-Dou and Miguel Gallegos and Kyunghoon Han and Florian N. Brünig and Joshua T. Berryman and Alexandre Tkatchenko},
      year={2026},
}

@misc{pytorch,
  title        = {PyTorch},
  author       = {{PyTorch Foundation}},
  howpublished = {\url{https://pytorch.org/}},
  year         = {2026},
  note         = {Accessed: 2026-02-19}
}

@misc{pytorchgeometric,
  title        = {PyTorch Geometric},
  author       = {{PyG Team}},
  howpublished = {\url{https://pyg.org/}},
  year         = {2026},
  note         = {Accessed: 2026-02-19}
}

@misc{jax2018github,
  title        = {{JAX}: composable transformations of {P}ython+{N}um{P}y programs},
  author       = {Bradbury, James and Frostig, Roy and Hawkins, Peter and Johnson, Matthew James and Katariya, Yash and Leary, Chris and Maclaurin, Dougal and Necula, George and Paszke, Adam and VanderPlas, Jake and Wanderman-Milne, Skye and Zhang, Qiao},
  howpublished = {\url{https://github.com/jax-ml/jax}},
  year         = {2018}
}

@misc{so3lr_github,
  title        = {{SO3LR}: {SO3krates} and Universal Pairwise Force Field for Molecular Simulation},
  author       = {{general-molecular-simulations}},
  howpublished = {\url{https://github.com/general-molecular-simulations/so3lr}},
  year         = {2025},
  note         = {GitHub repository}
}

@article{Rossi2013Impact,
  author    = {Rossi, Mariana and Scheffler, Matthias and Blum, Volker},
  title     = {Impact of Vibrational Entropy on the Stability of Unsolvated Peptide Helices with Increasing Length},
  journal   = {The Journal of Physical Chemistry B},
  year      = {2013},
  volume    = {117},
  number    = {18},
  pages     = {5574--5584},
  publisher = {American Chemical Society},
  issn      = {1520-6106}
}

@article{AlexPRL2011,
  title = {Unraveling the Stability of Polypeptide Helices: Critical Role of van der Waals Interactions},
  author = {Tkatchenko, Alexandre and Rossi, Mariana and Blum, Volker and Ireta, Joel and Scheffler, Matthias},
  journal = {Phys. Rev. Lett.},
  volume = {106},
  issue = {11},
  pages = {118102},
  numpages = {4},
  year = {2011},
  month = {Mar},
  publisher = {American Physical Society},
}

@article{AtomicInteractions,
  title     = {Analyzing Atomic Interactions in Molecules as Learned by Neural Networks},
  author    = {Esders, Malte and Schnake, Thomas and Lederer, Jonas and Kabylda, Adil and Montavon, Gr{\'e}goire and Tkatchenko, Alexandre and M{\"u}ller, Klaus-Robert},
  journal   = {Journal of Chemical Theory and Computation},
  year      = {2025},
  month     = jan,
  volume    = {21},
  number    = {2},
  pages     = {714--729},
  publisher = {American Chemical Society},
  issn      = {1549-9618}
}

@inproceedings{pairnorm2020,
  title={PairNorm: Tackling Oversmoothing in GNNs},
  author={Zhao, Lingxiao and Akoglu, Leman},
  booktitle={International Conference on Learning Representations},
  year         = {2020},
}

@inproceedings{dropedge2020,
  title={DropEdge: Towards Deep Graph Convolutional Networks on Node Classification},
  author={Rong, Yu and Huang, Wenbing and Xu, Tingyang and Huang, Junzhou},
  booktitle={International Conference on Learning Representations},
  year         = {2020},
}

@inproceedings{demystify_gat_oversmooth2023,
  author       = {Xinyi Wu and
                  Amir Ajorlou and
                  Zihui Wu and
                  Ali Jadbabaie},
  editor       = {Alice Oh and
                  Tristan Naumann and
                  Amir Globerson and
                  Kate Saenko and
                  Moritz Hardt and
                  Sergey Levine},
  title        = {Demystifying Oversmoothing in Attention-Based Graph Neural Networks},
  booktitle    = {Advances in Neural Information Processing Systems 36: Annual Conference
                  on Neural Information Processing Systems 2023, NeurIPS 2023, New Orleans,
                  LA, USA, December 10 - 16, 2023},
  year         = {2023},
}

@article{blum2009fhi,
  title   = {Ab initio molecular simulations with numeric atom-centered orbitals},
  author  = {Blum, Volker and Gehrke, Ralf and Hanke, Felix and Havu, Paula
             and Havu, Ville and Ren, Xinguo and Reuter, Karsten and Scheffler, Matthias},
  journal = {Computer Physics Communications},
  volume  = {180},
  number  = {11},
  pages   = {2175--2196},
  year    = {2009},
}

@article{huber1964robust,
  title   = {Robust Estimation of a Location Parameter},
  author  = {Huber, Peter J.},
  journal = {The Annals of Mathematical Statistics},
  volume  = {35},
  number  = {1},
  pages   = {73--101},
  year    = {1964},
}

\section*{Acknowledgements}
The authors express their gratitude to Tobias Henkes for the SO3krates Torch implementation. 
The authors also thank Miguel Gallegos for valuable discussions and feedback.
This work was supported by the Spanish Research Agency (AEI) under project  PID2024-161868OB-I00 [C3DRUM] funded by MCIN/AEI/10.13039/501100011033 and FEDER, PID2021-122464NB-I00, TED2021-129593B-I00, CNS2023-144561, PID2024-155562NB-I00 and Maria de Maeztu CEX2021-001202-M. E. R. also acknowledges the Generalitat de Catalunya for an ICREA Academia grant. 
S.S-D. and A.T. acknowledge support from the Luxembourg National Research Fund under grant FNR-CORE MBD-in-BMD (18093472) and the European Research Council under ERC-AdG grant FITMOL (101054629). The FHI-aims computations were performed on the Luxembourg national supercomputer MeluXina.

\section*{Author Contributions}
\textbf{À. S.}: Conceptualisation, Methodology, Software, Investigation, Validation, Writing - review \& editing.
\textbf{S. S-D.}: Conceptualisation, Methodology, Investigation, Validation, Writing - review \& editing.
\textbf{A. M-M.}: Conceptualisation, Supervision, Writing - review \& editing.
\textbf{S. G-C.}: Supervision, Resources, Writing - review \& editing, Funding acquisition.
\textbf{E. R.}: Supervision, Resources, Writing - review \& editing, Funding acquisition.
\textbf{A. T.}: Conceptualisation, Supervision, Resources, Writing - review \& editing, Funding acquisition.
\textbf{J. R-H.}: Conceptualisation, Supervision, Resources, Writing - review \& editing, Funding acquisition.

\section*{Competing Interests}
The authors declare no competing interests.

\end{document}


\title[Supplementary Information]{Supplementary Information: Machine Learning Multiscale Interactions}

\author[1,2,3]{\fnm{Àlex} \sur{Solé}}\email{jaume.alexandre.sole@upc.edu}

\author[3]{\fnm{Sergio} \sur{Suárez-Dou}}\email{sergio.suarezdou@uni.lu}

\author[1]{\fnm{Albert} \sur{Mosella-Montoro}}\email{albert.mosella@upc.edu}

\author*[2]{\fnm{Silvia} \sur{Gómez-Coca}}\email{silvia.gomez@qi.ub.edu}

\author*[2]{\fnm{Eliseo} \sur{Ruiz}}\email{eliseo.ruiz@qi.ub.edu}

\author*[3]{\fnm{Alexandre} \sur{Tkatchenko}}\email{alexandre.tkatchenko@uni.lu}

\author*[1]{\fnm{Javier} \sur{Ruiz-Hidalgo}}\email{j.ruiz@upc.edu}

\affil[1]{\orgdiv{Image Processing Group -- Signal Theory and Communications Department}, \orgname{Universitat Politècnica de Catalunya}, \orgaddress{\city{Barcelona}, \country{Spain}}}

\affil[2]{\orgdiv{Inorganic and Organic Chemistry Department and Institute of Theoretical and Computational Chemistry}, \orgname{Universitat de Barcelona}, \orgaddress{\city{Barcelona}, \country{Spain}}}

\affil[3]{\orgdiv{Department of Physics and Materials Science}, \orgname{University of Luxembourg}, \orgaddress{\city{Luxembourg City}, \country{Luxembourg}}}

\maketitle

\section{Experimental Details}
\label{sec:hyperparams}

Experiments were conducted on four different machines. Two machines were equipped with 8 NVIDIA RTX 3090 GPUs (24 GB memory), 2 AMD EPYC 7313 16-core CPUs, and 256 GB RAM. The third machine was equipped with 3 NVIDIA RTX 4090 GPUs (24 GB memory), 2 AMD EPYC 9474F 48-core CPUs, and 384 GB RAM. The fourth machine was equipped with 1 NVIDIA RTX 4090 GPU (24 GB memory), an AMD Ryzen 7 7700X 8-core CPU, and 64GB of RAM.  All implementations use PyTorch v2.4.0~\cite{pytorch}, PyTorch Geometric v2.6.1~\cite{pytorchgeometric} and ASE 3.26.0~\cite{ase}.

We train the model with a combined objective that penalises errors in both energies and forces using the smooth L1 loss, a scaled form of the Huber loss~\cite{huber1964robust}. For a scalar residual $x$, this loss is defined as
\begin{equation*}
\ell_{\mathrm{SL1},\delta}(x)
=
\begin{cases}
\frac{1}{2\delta}x^2, & |x| < \delta,\\
|x| - \frac{\delta}{2}, & \mathrm{otherwise},
\end{cases}
\end{equation*}
where $\delta$ is the transition point between the quadratic and linear regimes. For a configuration with reference energy $E$ and reference forces $\mathbf{F}_i \in \mathbb{R}^{3}$, the per-configuration loss is
\begin{equation}
\mathcal{L}
=
(1-\beta)\,\ell_{\mathrm{SL1},\delta}\bigl(E - \hat{E}\bigr)
+
\frac{\beta}{3N}\sum_{i=1}^{N}\sum_{\alpha=1}^{3}
\ell_{\mathrm{SL1},\delta}\bigl(F_{i\alpha} - \hat{F}_{i\alpha}\bigr),
\label{eq:loss_energy_forces}
\end{equation}
where $\beta \in [0,1]$ controls the relative weighting between energy and force supervision, and the factor $3N$ normalises the force term by the number of force components in the system. For all the experiments we used $\beta = 0.99$.

All the SO3krates models are trained using 132 feature dimensions, 4 heads, 1,2,3 degrees, a radius cutoff of $r_c$ = 5.0 \AA, 3 layers, and Ziegler–Biersack–Littmark universal repulsive potential~\cite{ZBLrepulsion}. For the SO3krates-MuSE model, we used the same base configuration but with $s=3$ scales, $\gamma$ = 1.5 to increase the radius at each scale, and $\varphi$ = 0.5, which halved the number of nodes at each scale. The $r_{pool}$ was set to 7.5 \AA for the first pooling layer and 11.25 \AA for the second pooling layer.

For all the MD22~\cite{MD22} and TEA2023~\cite{crashtest} compounds, we followed the methodology presented in SO3krates~\cite{so3krates2}, where all the compound are trained 1M steps, using AMSGrad optimizer~\cite{amsgrad} with a 5000-step linear warmup to a learning rate of $10^{-3}$ followed by an exponential decay of 0.7 every 100k steps. The batch size was 10 for all compounds except the double-walled nanotube, which has a batch size of 2. For the MD22 the training samples are reported, for validation we used 500 samples and the rest for test. For the TEA2023, we used the given splits

For the GEMS~\cite{GEMS} alanine subset, we followed the methdology presented by SpookyNet~\cite{unke2021spookynet} and we trained all the models using AMSGrad optimizer~\cite{amsgrad} with a learning rate of $10^{-3}$, every 1k steps is evaluated the valdiation set and if the validation loss does not decrease for 25 consecutive evaluations, the learning rate is decrased by 0.5. The training stops when the learning rate drops below $10^{-5}$.
All SO3LR experiments were carried out with the JAX~\cite{jax2018github} implementation available in the SO3LR repository~\cite{so3lr_github}.

For the model agnostic comparison experiments, we used PaiNN models with 128 feature dimensions, 3 interaction blocks, 32 radial basis functions, and a radius cutoff of $r_c$ = 5.0 \AA, and MACE models with a radius cutoff of $r_c$ = 5.0 \AA, 8 radial basis functions, polynomial cutoff order 3, maximum angular degree $\ell_{\max}=2$, hidden irreps $256\times 0e + 256\times 1o + 256\times 2e$, correlation order 3, and 2 interaction blocks. For the PaiNN-MuSE and MACE-MuSE models, we used the same base configurations but with $s=3$ scales, $\gamma$ = 1.5 to increase the radius at each scale, and $\varphi$ = 0.5, which halved the number of nodes at each scale. The $r_{pool}$ was set to 7.5 \AA for the first pooling layer and 11.25 \AA for the second pooling layer.

\section{Comparison with other long-range approaches}

\begin{figure}[p]
    \centering
    \includegraphics[width=\textwidth,height=0.868\textheight,keepaspectratio]{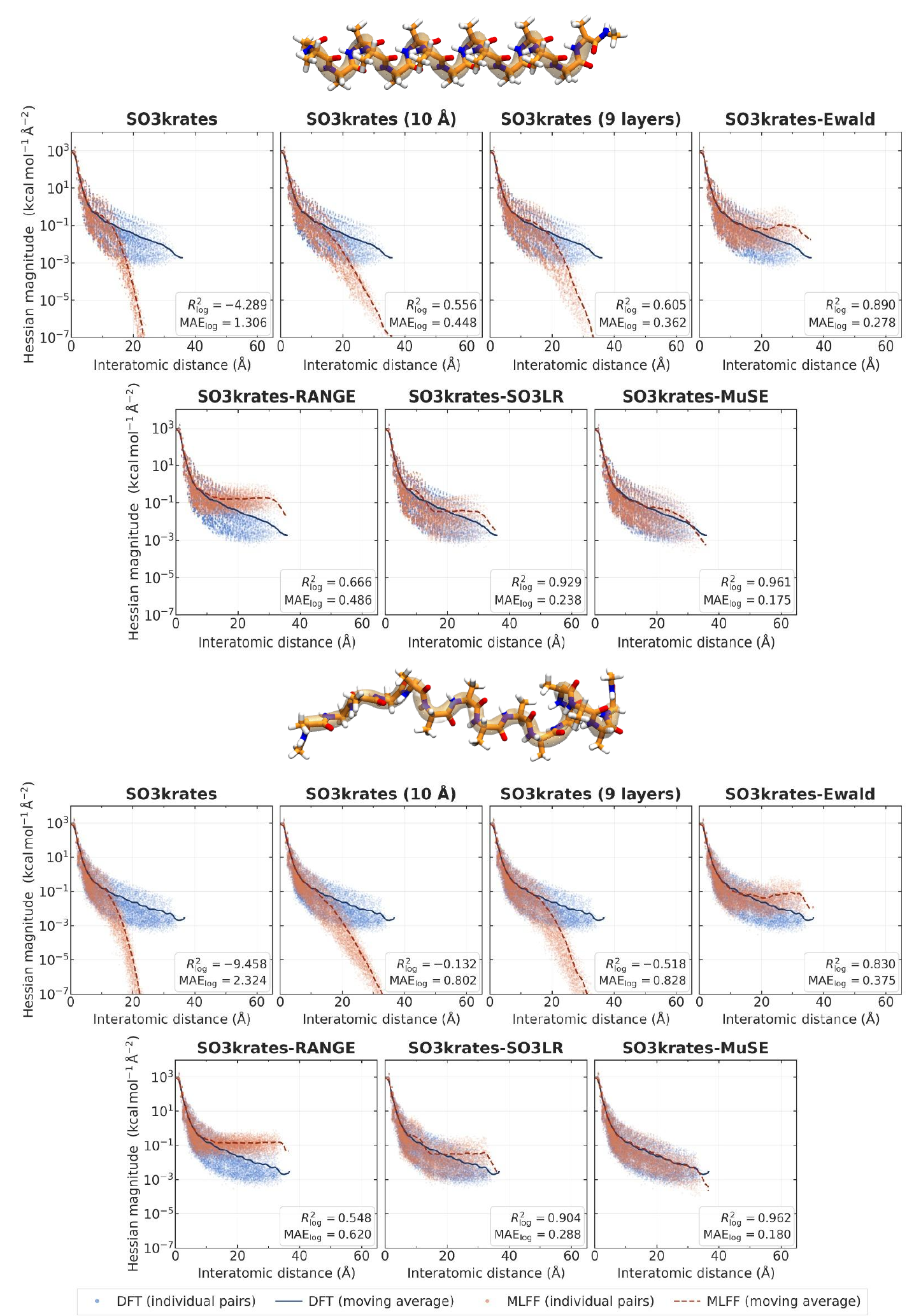}
    \caption{Hessian interaction--distance analysis for the two additional Ace-Ala$_{15}$-NMe conformers not shown in the main text: a $3_{10}$-helical conformer (top) and a nearly unfolded conformer (bottom). For each model, individual pairwise Hessian magnitudes are plotted against interatomic distance and compared with the corresponding DFT reference. Blue markers and solid lines denote DFT individual pairs and their moving average, respectively, whereas orange markers and dashed lines denote the corresponding MLFF predictions. Insets report the $\log_{10}$-space metrics used throughout the paper, namely $R^2_{\log}$ and $\mathrm{MAE}_{\log}$.}
    \label{fig:hessian_methods_suppl}
\end{figure}

For completeness, Fig.~\ref{fig:hessian_methods_suppl} reports the same Hessian interaction--distance analysis for the two remaining Ace-Ala$_{15}$-NMe conformers considered in this work, namely the $3_{10}$-helical and the nearly unfolded structures. Consistent with the trends observed in the main text, the baseline SO3krates model shows a marked deterioration in the recovery of distance-dependent couplings, particularly beyond the local regime, whereas simply increasing the cutoff or network depth yields only partial and conformer-dependent improvements. By contrast, methods explicitly designed to enhance non-local reasoning recover the DFT Hessian profile more faithfully across intermediate and long distances. Among the approaches compared, SO3krates-MuSE again provides the closest overall agreement with the DFT reference, reproducing both the decay and the magnitude of the moving-average Hessian profile with the highest consistency across the two conformers. This behaviour is particularly notable for the nearly unfolded structure, where the extended geometry makes the description of medium- and long-range intramolecular couplings more demanding. Taken together with the results for the $\alpha$-helical and unfolded conformers shown in the main text, these additional examples further support the conclusion that the multiscale hierarchy improves the recovery of non-local interactions without requiring explicit Hessian supervision.  

\section{MuSE scale ablation}
\label{subsec:muse_scale_ablation}

\begin{figure}[!htb]
    \centering
    \includegraphics[width=\linewidth]{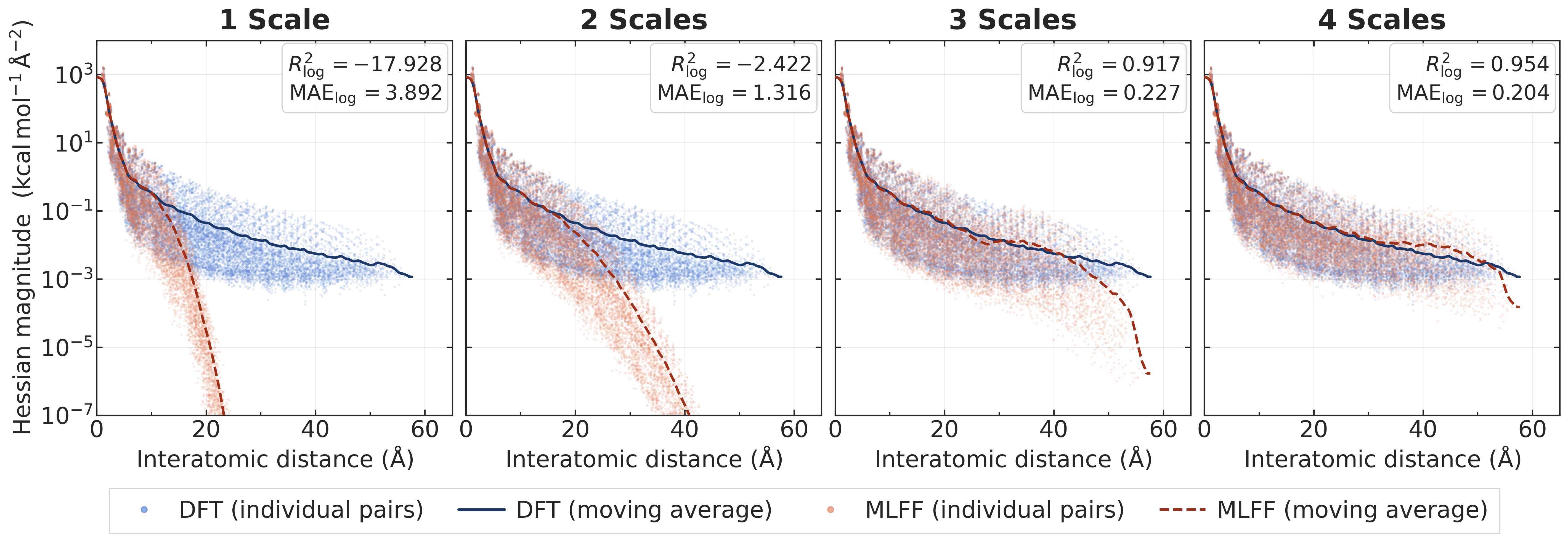}
    \caption{MuSE scale ablation on the Hessian interaction--distance profile for the unfolded Ace-Ala$_{15}$-NMe conformer. Each panel compares DFT reference Hessian magnitudes (blue) with MLFF predictions (orange) for MuSE variants with one to four scales. Scatter points denote individual pairwise Hessian magnitudes. Solid and dashed curves denote the corresponding moving averages for DFT and MLFF, respectively. Insets report the $\log_{10}$-space agreement metrics, $R^2_{\log}$ and $\mathrm{MAE}_{\log}$.}
    \label{fig:scale_ablations_ala15_unfolded}
\end{figure}

We performed a scale ablation study to examine how the Hessian profile changes with hierarchy depth in MuSE. The SO3krates backbone and the MuSE construction were kept fixed. The number of scales was then varied from one to four. Fig.~\ref{fig:scale_ablations_ala15_unfolded} shows the resulting Hessian interaction--distance profiles for the unfolded Ace-Ala$_{15}$-NMe conformer. With one scale, MuSE is effectively atomistic and does not contain a coarse hierarchy. This variant fails to reproduce the long-range Hessian tail. Its moving average decays several orders of magnitude too rapidly and gives a strongly negative $R^2_{\log}$. Introducing a second scale improves the intermediate-distance regime. However, the predicted profile still collapses at longer separations and remains poorly correlated with the DFT reference.

The fully unfolded conformer is the most demanding case in this comparison because it contains the largest pairwise separations. For conformers that have already started to fold, the Hessian profile is already reproduced with high fidelity by the three-scale model, as shown for the nearly unfolded structure in Fig.~\ref{fig:hessian_methods_suppl}. The remaining sensitivity to the number of scales is therefore concentrated in the fully extended geometry.

The main improvement appears when the hierarchy contains three scales. In this setting, the MLFF moving average follows the DFT trend over most of the accessible distance range. The corresponding metrics are $R^2_{\log}=0.917$ and $\mathrm{MAE}_{\log}=0.227$. This result indicates that the third scale provides a sufficiently coarse communication pathway for the extended Ala$_{15}$ conformation. At the same time, the atomistic branch remains available to resolve local chemical detail. Increasing the hierarchy to four scales yields a further but comparatively modest improvement. The metrics improve to $R^2_{\log}=0.954$ and $\mathrm{MAE}_{\log}=0.204$. Taken together with the conformers in Fig.~\ref{fig:hessian_methods_suppl}, these results identify three scales as the most favourable trade-off for this system. In most conformations, the Hessian profile is already essentially indistinguishable from the DFT reference. The fourth scale mainly reduces a small residual error in the fully stretched conformer.

\begin{figure}[!htb]
    \centering
    \includegraphics[width=\linewidth]{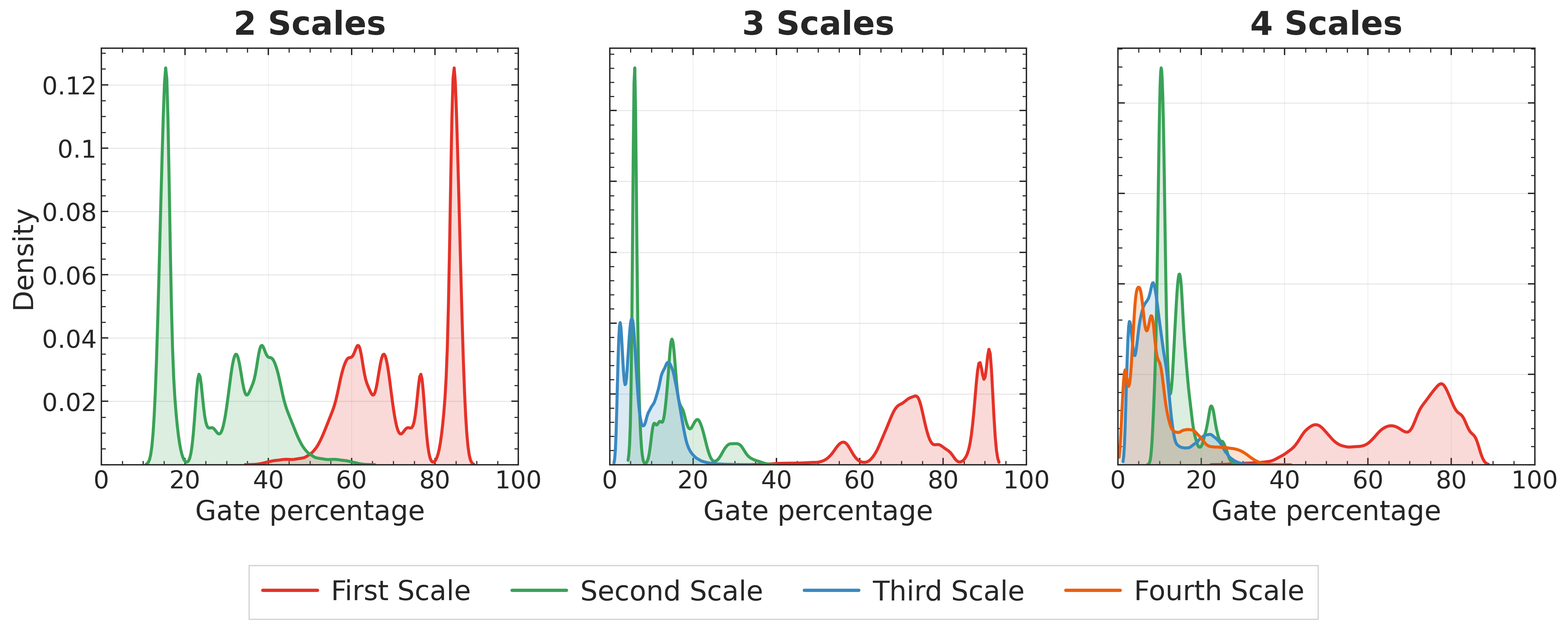}
    \caption{Decoder gate distributions for MuSE models with two, three, and four scales on the unfolded Ace-Ala$_{15}$-NMe conformer. Each panel shows the density of gate percentages for the available scales.}
    \label{fig:gate_distributions_by_method}
\end{figure}

We also analysed the gate coefficients produced by the MuSE decoder (Fig.~\ref{fig:gate_distributions_by_method}). In MuSE, each scale-specific representation is first upsampled to atomistic resolution. The gates are atom-wise routing coefficients used by the decoder to combine scale-specific embeddings into the final atom-wise representation before energy prediction. They quantify how information from each scale is routed into the final descriptive embedding. They should not be interpreted as direct scale-wise energy contributions. The total energy is predicted only after the gated multiscale embedding has been formed. Forces are then obtained by differentiating that energy with respect to the atomic coordinates.

The gate distributions provide a descriptive view of the decoder routing. For the two-scale model, the gate of the first scale is concentrated at high percentages, mainly between 55\% and 90\%. The gate of the second scale occupies lower and intermediate percentages, with most density between 10\% and 45\%. For the three-scale model, the gate of the first scale again carries the largest percentages for many atoms, with density extending from about 55\% to above 90\%. The gates of the second and third scales are mainly assigned lower percentages. Their distributions are concentrated below approximately 25\%, with local maxima at small gate values. For the four-scale model, the routing is more dispersed. The gate of the first scale spans both low and high percentages, whereas the gates of the second, third, and fourth scales are mostly concentrated below approximately 25\%. These distributions show how the decoder allocates representation-space weight across the available scales for each model. They are not used here to assign direct energetic contributions to individual scales.

\bibliography{refs}